\documentclass[twocolumn]{aastex631}

\usepackage{amsmath}

\begin{document}

\title{Formation process of young stellar population in Messier 16 from a kinematic perspective}

\author[0000-0001-5797-9828]{Beomdu Lim}
\affiliation{Department of Earth Science Education, Kongju National University, 
56 Gongjudaehak-ro, Gongju-si, Chungcheongnam-do 32588, Republic of Korea}
\affiliation{Earth Environment Research Center, Kongju National University, 56 Gongjudaehak-ro, Gongju-si, Chungcheongnam-do 32588, Republic of Korea}
\affiliation{Korea Astronomy and Space Science Institute, 776 
Daedeok-daero, Yuseong-gu, Daejeon 34055, Republic of Korea}

\author[0000-0001-6842-1555]{Hyeong-Sik Yun}
\affiliation{Korea Astronomy and Space Science Institute, 
776 Daedeok-daero, Yuseong-gu, Daejeon 34055, Republic of Korea}

\author[0000-0002-1609-3956]{Hyun-Jeong Kim}
\affiliation{Earth Environment Research Center, Kongju National University, 56 Gongjudaehak-ro, Gongju-si, Chungcheongnam-do 32588, Republic of Korea}
\affiliation{Korea Astronomy and Space Science Institute, 776 
Daedeok-daero, Yuseong-gu, Daejeon 34055, Republic of Korea}

\author[0009-0005-4645-0263]{Yuna Lee}
\affiliation{Department of Earth Science Education, Kongju National University, 56 Gongjudaehak-ro, Gongju-si, Chungcheongnam-do 32588, Republic of Korea}

\author[0000-0001-8969-0009]{Jae-Rim Koo}
\affiliation{Earth Environment Research Center, Kongju National University, 56 Gongjudaehak-ro, Gongju-si, Chungcheongnam-do 32588, Republic of Korea}

\author[0000-0002-5097-8707]{Jongsuk Hong}
\affiliation{Korea Astronomy and Space Science Institute, 
776 Daedeok-daero, Yuseong-gu, Daejeon 34055, Republic of Korea}

\author[0000-0002-0418-5335]{Heeyoung Oh}
\affiliation{Korea Astronomy and Space Science Institute, 
776 Daedeok-daero, Yuseong-gu, Daejeon 34055, Republic of Korea}



\begin{abstract}
We present a kinematic study of young stars in Messier 16 (M16)
using the Gaia Data Release 3 and high-resolution spectra. 
A total of 345 stars are selected as genuine members using 
the published lists of X-ray, infrared sources, and early-type stars 
as well as the Gaia data. There is severe differential 
reddening across this region and the reddening law of the 
intracluster medium appears abnormal. The distance to M16, 
derived from the parallaxes of the members, is about 1.7 kpc. 
The ages of members, estimated by comparing their color-magnitude 
diagram with theoretical isochrones, range from 1 Myr to 4 Myr. 
This star-forming region is composed of an open cluster 
(NGC 6611) and a distributed population. This cluster shows 
a clear pattern of expansion and rotation. Some of the distributed 
population are spatially associated with the gas pillars located 
at the ridge of H {\scriptsize II} bubble. In particular, several 
stars moving away from the cluster are physically associated 
with the northeastern pillar. In addition, their younger ages 
support the idea that the formation of these stars was triggered 
by the feedback from massive stars in NGC 
6611. On the other hand, the other stars do not show systematic 
radial or stream motions; therefore, they likely formed through 
spontaneous star formation events. We discuss 
the formation of young stars in the context of cluster expansion, 
spontaneous star formation, and feedback-driven star 
formation, and suggest that all of these mechanisms possibly 
contributed to their formation.
\end{abstract}



\section{Introduction} \label{sec1}
Stellar associations are large stellar systems composed 
of young stars spread out over a scale of tens of parsecs. In 
particular, OB associations are interesting stellar 
systems because they are the sites of cluster formation 
as well as massive star-forming regions. One or more stellar 
clusters, stellar groups, and a distributed population 
constitute OB associations \citep[and references therein]
{B64,LHL23}. Furthermore, OB associations form larger star 
formation complexes at spatial 
scales of hundreds of parsecs \citep{G18}. These facts indicate 
that star formation taking place at different spatial scales 
shows self-similar patterns \citep{EEPO0}. OB associations 
are therefore ideal targets for understanding star formation 
processes from small to large scales.

Theoretical studies have proposed three different models 
to explain the formation process of OB associations. One 
model postulates that early dynamical evolution of embedded 
stellar clusters can result in the formation of OB associations 
\citep{T78,H80,LMD84,KAH01,BK13,BK15}. In the model, rapid 
gas expulsion promotes the expansion of the infant clusters. 
In addition, stellar feedback is also a potential source 
of influence on the early dynamics of infant star clusters 
before gas expulsion \citep{GBR17}. A number of observational 
studies have actually found the patterns of cluster expansion 
\citep[and references therein]{CJW19,KHS19,DDL24}.

It is also possible that the internal structure of 
OB associations has been inherited from the filamentary 
structure of their natal clouds. The images observed 
at submillimeter wavelengths have shown that a number 
of molecular clouds have filaments \citep{A15}. Either 
turbulence or magnetic fields are likely involved in 
the formation of the filaments \citep[etc]{L81,PJG01,
WLE19,DHF20}. Several theoretical studies have 
successfully simulated star formation along the filaments 
of molecular clouds \citep{BBV03,BSC11,FOG24}. Stellar 
clusters form naturally in high-density regions with 
high star formation efficiency, while low star 
formation efficiency is responsible for the formation of 
sparse groups \citep{BSC11,K12}. 

Another possible process associated with the formation 
of OB associations is self-regulating star formation by 
feedback from massive stars \citep{EL77}. Ionization and 
shock fronts generated by far-ultraviolet radiation from 
massive stars proceed into the remaining clouds and compress 
them to form a new generation of stars. While several previous 
studies have raised questions about this model 
\citep{DEB12,DEB13,DHB15,YLL18,YLK21}, the signposts 
of feedback-driven star formation have steadily reported 
from many observations \citep{FHS02,SHB04,ZPD07,LSK14}.

We have performed a systematic survey of six OB 
associations in the Galaxy using the Gaia data \citep{gdr2,
gedr3} and high-resolution spectroscopy \citep{LNG19,LHY20,
LNH21,LNH22,LHL23}. The survey results are summarized by \citet{L24}. 
Stellar clusters appear to form from spontaneous 
star formation events, 80\% of which are undergoing 
expansion. This feature will further contribute to the 
expansion of OB associations in several million years. 
In some associations, smaller groups of stars 
were found around stellar clusters. Their origins seem 
to be related to either spontaneous star formation at 
their current locations or feedback-driven star formation. 
Cluster expansion, star formation in non-uniform molecular 
clouds, and feedback-driven star formation likely play 
roles in the formation of OB associations. However, further 
study is needed to reach more definitive conclusions about 
the formation process of stellar associations.

The star-forming region (SFR) Messier 16 (M16), which is 
a part of the Serpens OB1 association, is a useful laboratory 
to examine the proposed models for the formation of stellar 
associations. M16 is about 1.7 -- 2.1 kpc away from the Sun 
\citep{HMS93,BKP99,BSB06,GPM07,CJV18,KHS19}. Star formation is 
actively taking place in this region. \citet{HMS93} 
performed a comprehensive study of young stellar population in 
M16 using photometric and spectroscopic data. They identified 
a huge number of intermediate-mass members as well as 
early-type stars showing emission lines. About 340 young stellar 
objects (YSOs) with infrared (IR) excess emission were identified by 
\citet{GPM07}. \citet{IRW07} conducted an extensive imaging 
survey with {\it Spitzer} and found more than 400 YSOs. Later, 
\citet{GCM12} detected 1755 X-ray sources, 1183 of which are 
probable members. 

A large fraction of young stars are concentrated in the 
young open cluster NGC 6611 found at the center of M16 
\citep{IRW07}. In addition, there are several small groups 
of stars and a distributed population spread over this 
SFR. The so-called `Pillars of Creation' are located at 
the south-east ridge of the H {\scriptsize II} bubble 
in M16 (hereafter southern pillars). This gas structure 
has been thought to be the 
site of feedback-driven star formation as systematic 
age sequences of YSOs and starless cores along the 
pillars were reported \citep{FHS02,STN02}. However, 
\citet{IRW07} argued against the feedback-driven star 
formation in the pillars because they could not find 
significant evolutionary sequence from NGC 6611.

This study aims to understand the formation process 
of young stellar population in M16 from a kinematic 
perspective, which will help to draw a conclusion 
on the general formation process of stellar association. A 
description of our observational data is covered in Section~\ref{sec2}. 
We present the results of this study in Section~\ref{sec3} 
and discuss the formation of young stellar population 
in Section~\ref{sec4}. Finally, our results are summarized 
in Section~\ref{sec5}.

\section{Data} \label{sec2}

\subsection{Member selection}\label{sec21}
We targeted young stars within a $1^{\circ}\times 1^{\circ}$ 
region centered at R.A.$= 18^{\mathrm{h}} \ 18^{\mathrm{m}} \ 
48_{\cdot}^{\mathrm{s}}00$, decl.$= -13^{\circ} \ 48^{\prime} 
\ 24\farcs0$ (J2000). Since this SFR is very close to 
the Galactic plane ($b \sim 0\fdg793$), member selection 
is the most important procedure to reliably investigate 
this region. To select genuine members, it is necessary 
to understand the properties of young stars. Techniques for 
selecting members based on multi-wavelength data \citep{Gaia16} 
are well reviewed \citep{WKZ23}. As with our previous studies 
\citep{LHY20,LNH21,LNH22,LHL23}, member selection was performed 
in two steps.

First, we identified member candidates using publicly 
available data. Early-type stars are potential member candidates 
given their short lifetime. A total of 142 early stars (from O to A) 
spread over M16 were taken from the database of Morgan–Keenan 
classification compiled by \citet{Sk14}, with the most recent 
classification adopted in cases of duplication. 

To identify low-mass member candidates with a warm 
circumstellar disk, we obtained a list of IR sources 
from the GLIMPS survey \citep[]{CBM09,https://doi.org/10.26131/irsa405}. Only the 
photometric data with errors lower than 0.1 mag were used 
for the reliable classifications of YSOs. Normal main 
sequences stars, in general, have photospheric colors of 
about 0 in mid-infrared passbands. However, we found 
offsets of about 0.29 for the main sequence stars 
in the colors combined with 3.6 \micron 
\ or 4.5 \micron \ data. In addition, there is an 
offset of about 0.06 in the $[3.6] - [4.5]$ color. 
We corrected for the offsets in the associated colors.

\begin{figure}[t]
\epsscale{1.0}
\plotone{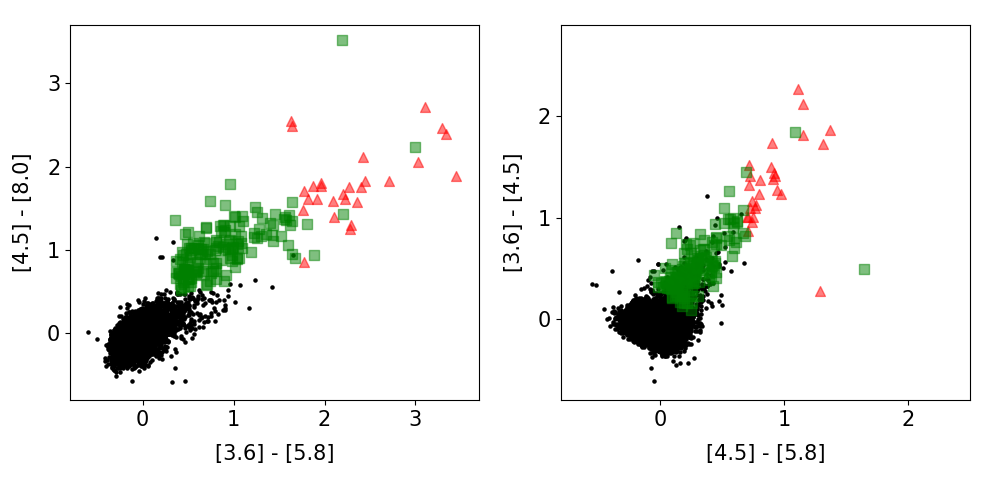}
\caption{Color-color diagrams of IR sources from the 
GLIMPS survey \citep{CBM09}. The red, green, and black 
dots represent Class 0/I, Class II, and the other sources 
without definite IR excess emission.}\label{fig1}
\end{figure}

Using the classification scheme of \citet{GMM08}, 
we identified YSOs after removing some contaminants 
such as active galactic nuclei and the emission 
from polyaromatic hydrocarbons. A total of 185 YSO 
candidates were found from 
the GLIMPS data, of which 27 and 158 sources were 
identified as Class 0/I and Class II objects, 
respectively. Figure~\ref{fig1} displays the 
color-color diagrams of the IR sources in M16. 

In addition, most YSOs and some early-type stars 
are X-ray sources \citep[etc]{FDM03,CMP08,
GCM12,RN16}. We obtained the most complete list of 
X-ray sources published by \citet{BGP13}, which 
contains a total of 2830 X-ray sources in M16. 

The catalogs of early-type stars, YSO candidates, 
and X-ray sources were then matched with the data 
from the Gaia Data Release 3 \citep{gdr3} with a 
searching radius of 1$^{\prime\prime}$. All 
counterparts of the early-type stars were identified 
in the Gaia data. We found the counterparts 
of 91 YSOs (3 Class 0/I and 88 Class II) and 1402 
X-ray sources through cross-identification with 
the Gaia sources, creating a catalog of 59,294 stars 
distributed across M16. The zero-point offsets for 
the parallaxes of these stars were corrected by 
following the recipe in \citet{LBB21}.

\begin{figure}[t]
\epsscale{1.0}
\plotone{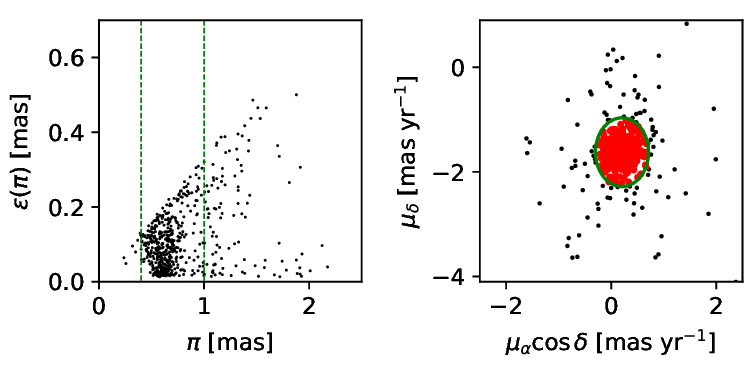}
\caption{Parallaxes (left) and PMs (right) 
of member candidates. The left panel shows 
the distribution of the parallaxes and 
their associated errors for stars brighter 
than 18 mag in the $G_{\mathrm{RP}}$ band. 
We searched for the genuine members between 
1.0 and 2.5 kpc (dashed lines). The right panel 
displays the PMs of member candidates distributed 
within the upper and lower bounds of the distance. 
The ellipse (green) shows the region confined
within three times the standard deviation from 
the weighted mean PMs, where the inverse of the 
squared PM error is used as the weight value. The 
red dots represent the most probable members in 
M16.}\label{fig2}
\end{figure}

\begin{figure}[t]
\epsscale{1.0}
\plotone{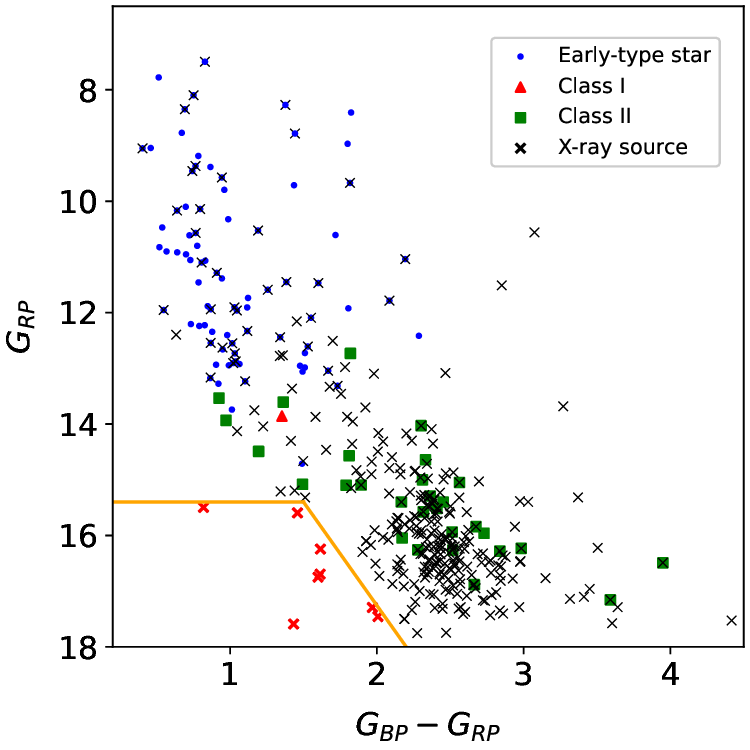}
\caption{CMD of members. Several 
member candidates (red crosses) below the orange 
solid lines are excluded from the final member list, while 
the other candidates are considered as genuine members.}\label{fig3}
\end{figure}

Figure~\ref{fig2} displays the distributions of the 
parallaxes and proper motions (PMs) of member candidates. 
Most candidates are found in a parallax 
range from 0.4 mas to 1.0 mas. Since most of them 
are found in narrow ranges of PMs (right panel), these 
stars are likely genuine members. We selected candidates 
within the three times the standard deviation from the mean 
PMs as members, where the mean and standard deviation were 
repeatedly computed until these values converged to constant 
values. 

This criterion excludes three O-type stars, BD-13 4923 
\citep[O4V((f)) + O7.5V;][]{SGE09}, 
BD-13 4927 \citep[O7II(f);][]{SMW11}, 
and HD 168076 \citep[O4IV(f);][]{MSA16}. 
The Renormalised Unit Weight Error (RUWE) of BD-13 4923 and 
HD 168076 are much greater than the expected value of 1.4 for 
a single star \citep{gdr3}. The high stellar density around 
these stars or their multiplicity may be responsible for the 
poor astrometric parameters. The other star 
BD-13 4927 has the good qualities of astrometric parameters 
(RUWE $=$ 1.056), but its PMs slightly deviate from the 
criterion we adopted. The distance to this star computed 
from the inverse of its Gaia parallax is about 1.65 kpc. 
Given the early spectral type and the distance, BD-13 4927 
is likely the genuine member of M16. We included the three 
early-type stars in our member list.

Figure~\ref{fig3} displays the color-magnitude diagram (CMD) 
of the selected members. There is a large spread in color 
at given magnitudes, implying a large amount of differential 
reddening across M16. The presence of the stars within 
the region outlined by two orange straight lines in the 
diagram cannot be explained by the stellar evolution of 
a coeval population and differential reddening. We excluded 
these stars from the member list. Finally, a total of 345 
stars were selected as the genuine members and are listed in 
Table~\ref{tab1}.

\begin{deluxetable}{ccccccccccccccccc}
\rotate
\tabletypesize{\tiny}
\tablewidth{0pt}
\tablecaption{List of members \label{tab1}}
\tablehead{
\colhead{R.A}. & \colhead{decl.} & \colhead{Source ID} & \colhead{$\pi$} & \colhead{$\epsilon(\pi)$} & \colhead{$\mu_{\alpha}\cos\delta$} & \colhead{$\epsilon(\mu_{\alpha}\cos\delta)$} & \colhead{$\mu_{\delta}$} & \colhead{$\epsilon(\mu_{\delta})$} & \colhead{RV$_{\mathrm{Helio}}$} & \colhead{RV$_{\mathrm{LSR}}$} &\colhead{$\epsilon$(RV)} & \colhead{$G$} & \colhead{$G_{BP}$} & \colhead{$G_{RP}$} & \colhead{$G_{BP} - G_{RP}$} & \colhead{Flag} \\
\colhead{[deg]} & \colhead{[deg]} & \colhead{-} & \colhead{[mas]} & \colhead{[mas]} & \colhead{[mas yr$^{-1}$]} & \colhead{[mas yr$^{-1}$]} & \colhead{[mas yr$^{-1}$]} & \colhead{[mas yr$^{-1}$]} & \colhead{[km s$^{-1}$]} & \colhead{[km s$^{-1}$]} & \colhead{[km s$^{-1}$]} & \colhead{[mag]} & \colhead{[mag]} & \colhead{[mag]}  & \colhead{-}  & \colhead{-}
}
\startdata
274.462682 & -13.848829 & 4146604939324925440 & 0.5782 & 0.0152 &   0.142 &  0.018 &  -1.638 &  0.014 &   -    &   -    &   -   &  9.354914 & 10.232370 &  8.407755 &  1.824615 & E   \\
274.532252 & -13.841184 & 4146604595727511296 & 0.6719 & 0.0480 &   0.148 &  0.057 &  -1.852 &  0.046 &  22.07 &  36.47 &  2.13 & 15.763083 & 16.973186 & 14.641831 &  2.331355 &  2  \\
274.534148 & -13.839977 & 4146604595727511040 & 0.5664 & 0.1042 &   0.205 &  0.121 &  -1.516 &  0.092 &   -    &   -    &   -   & 17.314571 & 18.546631 & 16.212688 &  2.333942 &   X \\
274.538686 & -13.781820 & 4146611497737061248 & 0.5825 & 0.0151 &   0.117 &  0.019 &  -1.239 &  0.014 &   -    &   -    &   -   & 12.334155 & 13.070502 & 11.469252 &  1.601251 & E X \\
274.551188 & -13.795520 & 4146611429017808128 & 0.5878 & 0.0407 &   0.159 &  0.046 &  -1.397 &  0.035 &   -    &   -    &   -   & 15.516960 & 16.379454 & 14.568625 &  1.810828 &  2  \\
\enddata
\tablecomments{Columns (1) and (2) : The equatorial coordinates of members (J2000). Column (3) : Source ID from \citet{gdr3}. Columns (4) and (5) : Absolute parallax and its standard error. Columns (6) and (7) : PM in the direction of right ascension and its standard error. Columns (8) and (9): PM in the direction of declination and its standard error. Columns (10 -- 12) : Heliocentric RV, RV in the local standard of rest frame, and its error. Columns (13--15): $G$ magnitude, $G_{BP}$ magnitude, and  $G_{RP}$ magnitude. Column (16) : $G_{BP} - G_{RP}$ color index. Column (17) : Classification of members. "E``, "1``, "2``, and "X`` represent early-type member, Class 0/1 YSO, Class II YSO, and X-ray source, respectively. The astrometric and photometric data were taken from the Gaia Data Release 3 \citep{gdr3}. This table is available in its entirety in machine-readable form.}
\end{deluxetable}

\subsection{Spectroscopic observations}\label{sec22}

\subsubsection{Optical spectroscopy}\label{sec221}
We performed multi-object spectroscopic observations 
of the YSO candidates on July 1 in 2020 and June 23 
in 2022 (UT) using the multi-object spectrograph Hectochelle \citep{SFC11} 
on the 6.5m telescope of the MMT observatory. The spectral 
resolution of Hectochelle $(R = \lambda / \Delta\lambda)$ 
is about 34000, however our observations were conducted to 
obtain high signal-to-noise ratios in 
a $2\times2$ binning mode. The spectra of 
118 YSO candidates were obtained with the RV31 filter in the 
spectral range of 5105 to 5300 \AA. We also allocated 
several tens of fibers to the blank sky to take sky 
spectra simultaneously. The exposure time for 
each frame is 35 minutes. At least three frames were 
taken for the same observation setup to remove cosmic rays 
and improve signal-to-noise ratios. Dome flat and ThAr lamp 
spectra were taken just before and after the target observations. 
Our observations are summarized in Table~\ref{tab2}.

The raw mosaic frames were reduced by using the IRAF\footnote{Image
Reduction and Analysis Facility is developed and distributed by the
National Optical Astronomy Observatories, which is operated by the
Association of Universities for Research in Astronomy under operative
agreement with the National Science Foundation.}/{\tt MSCRED}
packages following standard reduction procedures. We extracted 
one-dimensional spectra from the reduced frames 
using the {\tt dofiber} task in the IRAF/{\tt SPECRED} package. 
The target spectra were then divided by the dome-flat spectra 
to correct for pixel-to-pixel variation. Wavelength calibration 
was made with the ThAr spectra for both target and sky spectra. We 
subtracted sky spectra from the individual target spectra and 
then combined the sky-subtracted spectra into a single spectrum 
for the same target. Finally, all target spectra were normalized 
by using continuum levels traced from a cubic spline interpolation.

A total of 49 out of the 118 spectra were not analyzed 
because the spectra are dominated by continuum or have very low 
signal-to-noise ratios. We measured the radial velocities (RVs) 
of the other 69 YSO candidates from cross-correlation 
functions with several synthetic spectra using {\tt xcsao} task 
in the \textsc{RVSAO} package \citep{KM98}. We generated the 
synthetic spectra in a wide temperature range from 3500 to 10000 K 
for the solar abundance and $\log g = 4.0$ using {\tt SPECTRUM 
v2.76} \citep{GC94}\footnote{\url{http://www.appstate.edu/~grayro/spectrum/spectrum.html}} 
based on a grid of the ODFNEW model atmospheres \citep{CK03}. 
The RVs of given YSO candidates were obtained from the velocities 
at the strongest correlation peaks. The errors on RVs were estimated 
using the equation of \citet{KM98} as below:
\begin{equation}
\epsilon(\mathrm{RV}) = {3w \over 8(1+h/\sqrt{2}\sigma_a)}
\end{equation}
\noindent where $w$, $h$, and $\sigma_a$ represent the full widths 
at half-maximum of cross-correlation functions, their amplitudes, 
and the root mean square of antisymmetric components, respectively. 
\citet{TD79} defined the signal-to-noise ratio of a cross-correlation 
function as $r = h/\sqrt{2}\sigma_a$. This is a useful tool for 
selecting high-quality RV data. We excluded some RV measurements 
with $r$ values smaller than 6. The median error of the RVs used in 
this study is about 1.4 km s$^{-1}$. The RVs of YSO 
candidates were then converted to velocities in both the local 
standard of rest frame as well as the heliocentric frame 
using the \textsc{IRAF}/{\tt RVCORRECT} task. 

\begin{figure}[t]
\includegraphics[width=0.47\textwidth]{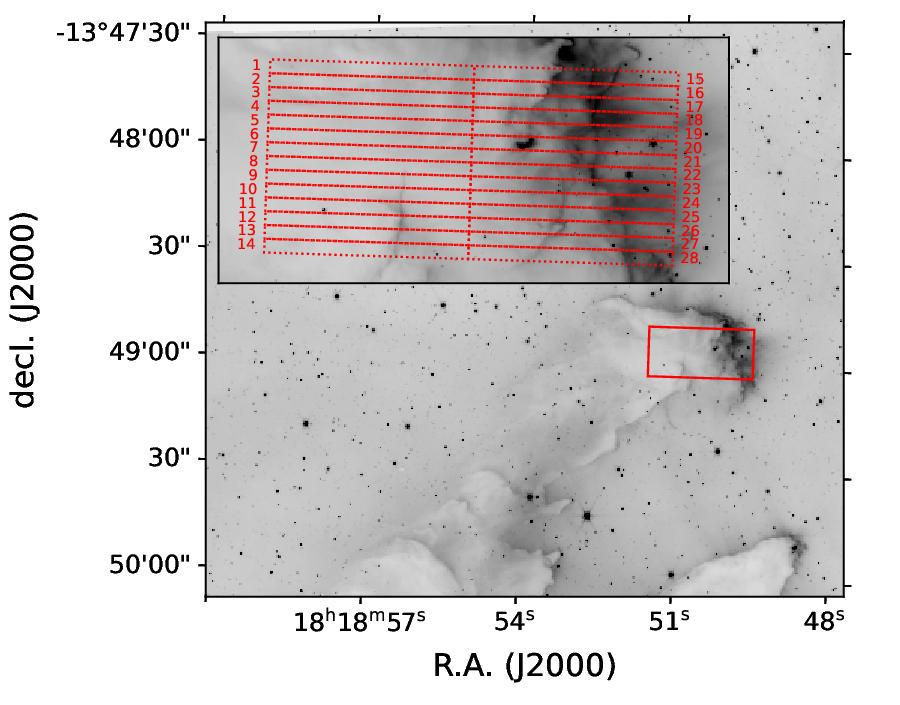}
\caption{IGRINS slit positions over one of the southern pillars. 
The gray scale images represent the NIRCam images obtained 
with the James Webb Space Telescope in F187N channel \citep{RKM23}. 
This image is made with publicly available data via the Mikulski Archive for Space 
Telescopes (MAST: \url{https://doi.org/10.17909/r5ne-bs21}). A total of 28 slits were 
used in our spectroscopic observations. 
The zoomed-in view shows the arrangement of the slits.}\label{fig4}
\end{figure}

\begin{deluxetable*}{ccccccc}
\tablewidth{0pt}
\tablecaption{Summary of spectroscopic observations \label{tab2}}
\tablehead{
\colhead{UT Date} & \colhead{Target}  & \colhead{Telescope$^{a}$} &
\colhead{Instrument} & \colhead{Spectral coverage} & \colhead{Integration time} & \colhead{Binning} \\
   \colhead{(YY-MM-DD)}     &  &  &  &  & \colhead{(s)} &  }
\startdata
 2020-07-01 & 67 YSOs & 6.5-m MMT & Hectochelle & 5105 -- 5300 \AA & $2100\times3$ & $2\times2$\\
 2022-06-23 & 51 YSOs & 6.5-m MMT & Hectochelle & 5105 -- 5300 \AA & $2100\times3$ & $2\times2$\\
 2016-07-19 & Pillar head & 2.7-m HJST & IGRINS & 1.45 -- 2.50 \micron & $300\times2$ & $1\times1$ \\
 2017-06-02 & Pillar head & 2.7-m HJST & IGRINS & 1.45 -- 2.50 \micron & $300\times2$ & $1\times1$   \\
 2017-06-03 & Pillar head & 2.7-m HJST & IGRINS & 1.45 -- 2.50 \micron & $300\times2$ & $1\times1$\\
 2017-06-04 & Pillar head & 2.7-m HJST & IGRINS & 1.45 -- 2.50 \micron & $300\times2$ & $1\times1$\\
 2017-06-06 & Pillar head & 2.7-m HJST & IGRINS & 1.45 -- 2.50 \micron & $300\times2$ & $1\times1$\\
 2024-04-16 & 2MASS J18190618-1344285 & 8.1-m GST & IGRINS & 1.45 -- 2.50 \micron & $325\times4$ & $1\times1$\\
 2024-04-17 & 2MASS J18190358-1345240 & 8.1-m GST & IGRINS & 1.45 -- 2.50 \micron & $300\times8$ & $1\times1$\\
 2024-04-18 & Cl* NGC 6611 GMD 521 & 8.1-m GST & IGRINS & 1.45 -- 2.50 \micron & $650\times4$ & $1\times1$\\
            & 2MASS J18190500-1343534 & 8.1-m GST &  
            IGRINS & 1.45 -- 2.50 \micron & $900\times4$ & $1\times1$\\
 2024-04-19 & 2MASS J18190315-1345111 & 8.1-m GST & IGRINS & 1.45 -- 2.50 \micron & $200\times4$ & $1\times1$\\
            & 2MASS J18190319-1345392 & 8.1-m GST &  
            IGRINS & 1.45 -- 2.50 \micron & $650\times1$ and $500\times3$ & $1\times1$\\         
\enddata
\tablenotetext{a}{MMT - Multiple Mirror Telescope, HJST - Harlan J. Smith Telescope, GST - Gemini South Telescope}
\end{deluxetable*}

\subsubsection{Near-infrared spectroscopy}\label{sec222}
The prominent features in M16 is the presence of the pillars 
exposed to the ultraviolet radiation from O-type stars 
in NGC 6611. We 
obtained the near-infrared spectra of the one of the southern 
pillars in the $H$ and $K$ bands using Immersion 
GRating INfrared Spectrometer (IGRINS -- \citealt{YJB10,PJY14,MSL18}) 
attached to the 2.7-m Harlan J. Smith Telescope at McDonald Observatory 
of the University of Texas at Austin on 2016 July 19, 2017 June 2, 
3, 4, and 6 (UT). The mapping observations were performed by placing 
a total of 28 slits across the gas pillar (Figure~\ref{fig4}). 
A nod technique, ON (source) – OFF (sky) – ON (source), was applied 
to the observing sequence to remove the sky background. The exposure 
time of each frame is 300 seconds. In addition, the observations 
of A0V stars were conducted to eliminate telluric absorption lines 
(Table~\ref{tab2}). 

We also obtained the spectra of six YSO members around 
the pillar located at the northeastern border of M16 
(hereafter northeastern pillar) and several A0V stars for 
telluric correction using the IGRINS installed on 
the Gemini South telescope (Table~\ref{tab2}). The spectroscopic 
observations were conducted through the Gemini Fast Turnaround 
program (GS-2024A-FT-204) from 2024 April 16 to 2024 April 19 (UT). 
The ABBA nodding sequence was applied to all observations. 

Data reduction was performed with the IGRINS Pipeline Package 
\citep{LGK17}. This pipeline conducts the basic procedures for 
data reduction, such as flat-fielding, bad pixel correction, wavelength 
calibration, background subtraction, and aperture extraction.
For the YSOs, the extracted one-dimensional spectra were divided 
by the one-dimensional spectra of A0 stars for telluric correction. 
For the mapping data, we stitched the two-dimensional spectra of 
the 28 slits and constructed the data cubes of the selected lines 
using the Python library PLOTSPEC \citep{KDO17}. Telluric lines were 
also corrected in the PLOTSPEC process by using the spectra 
of A0 stars. The emission lines detected and used for constructing 
data cubes are H$_2$ $1-0$ S(1), Br$\gamma$, 
He {\scriptsize I} 2.059 \micron, and [Fe {\scriptsize II}] 1.644 \micron.

The RVs of the observed YSOs were measured using the 
same technique and the models of stellar atmospheres as 
applied to the optical spectra \citep{KM98,GC94,CK03}.
To do this, we converted the wavelengths of the synthetic 
spectra to those in vacuum by using the relation of \citet{C96}. 
The near-IR spectrum of the YSO Cl* NGC 6611 GMD 521 is dominated 
by emission lines, and therefore we were unable to measure 
the RV of this YSO. The RVs of the other five YSOs were measured 
with a median error of about 2.0 km s$^{-1}$. 

The RVs of J18190319-1345392 were measured from both 
optical and near-infrared spectra. The difference 
between the RVs is about 1.1 km s$^{-1}$, 
which is similar to the typical error of RV measured 
in the optical spectra and smaller than the RV error in 
the near-infrared spectra. Therefore, it is unlikely that 
there is a significant offset between the two datasets 
(IGRINS and Hectochelle) in the RV.

\begin{figure*}[t]
\epsscale{1.0}
\plottwo{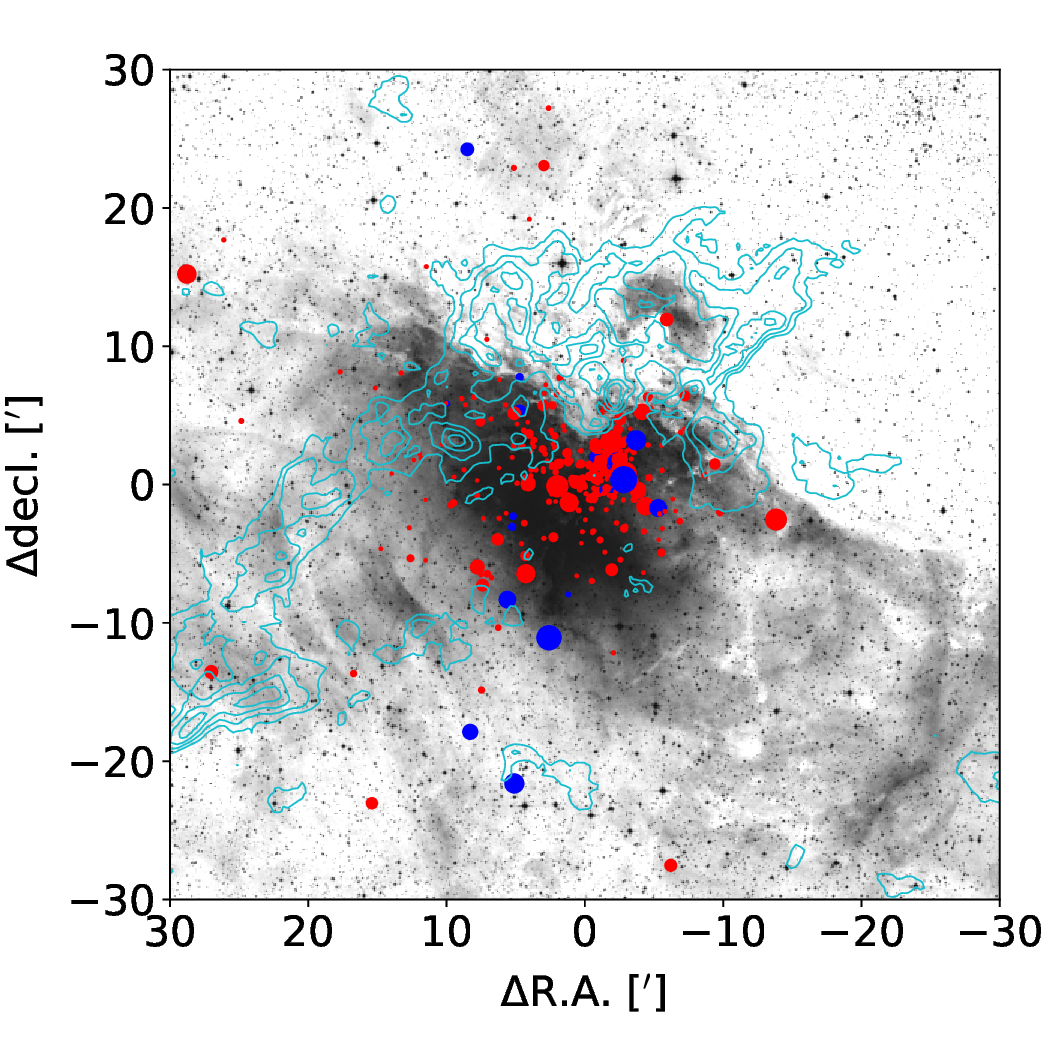}{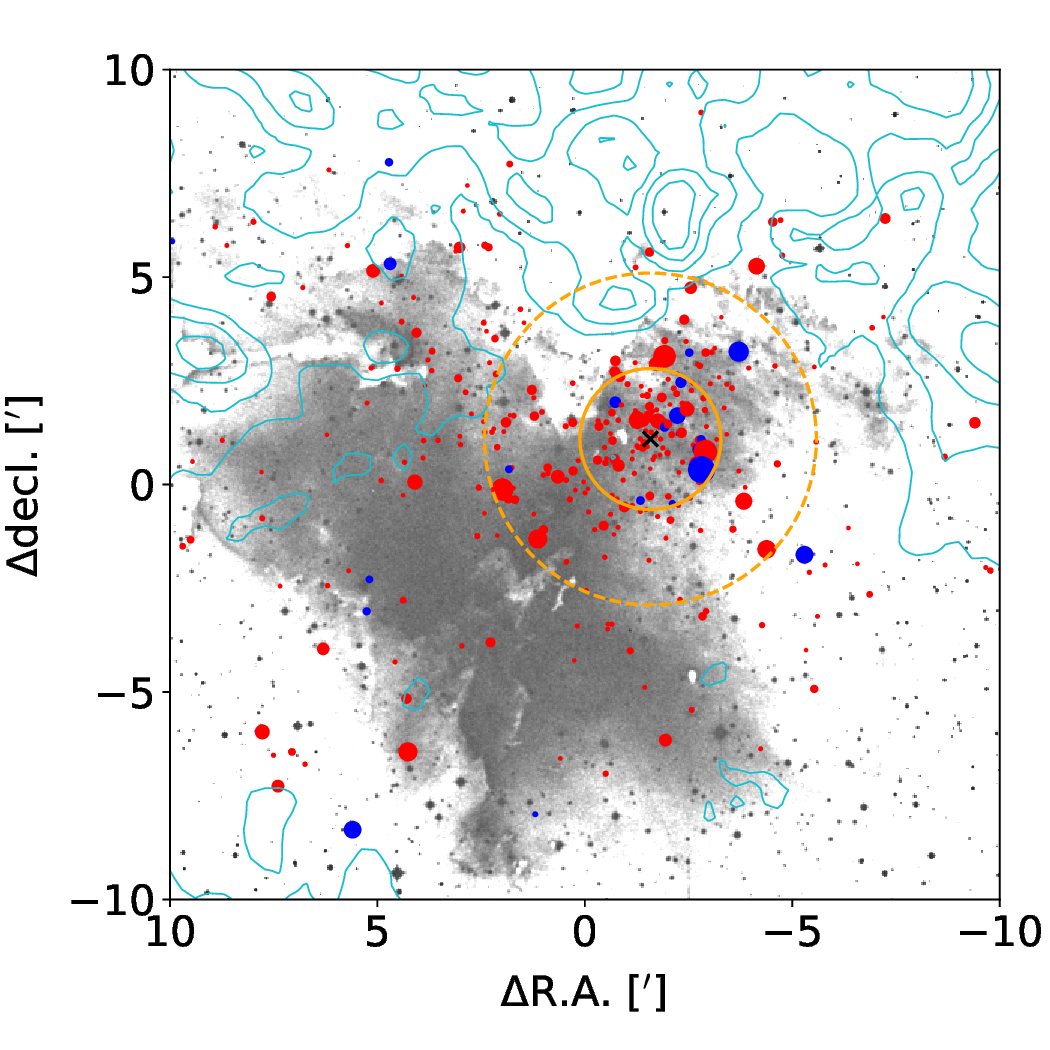}
\caption{Spatial distribution of members and gas. The background images were taken from the Digitized Sky Survey available at MAST \citep{https://doi.org/10.17909/T9QP4J}. The positions 
of stars with good astrometric parameters (RUWE $<$ 1.4 and duplicate 
flag $=$ 0) are shown by red dots, while the blue dots represent the 
stars with poor astrometric parameters. The size of dots is 
proportional to the brightness of individual 
stars. The cyan contours shows the distribution of molecular gas 
traced by $^{13}CO \ J = 1 - 0$ line. The contour levels are integrated 
intensities of 5, 10, 15, and 20~K~km~s$^{-1}$. The coordinates of stars and gas 
are relative to R.A.$= 18^{\mathrm{h}} 
\ 18^{\mathrm{m}} \ 48_{\cdot}^{\mathrm{s}}00$, 
decl.$= -13^{\circ} \ 48^{\prime} \ 24\farcs0$ (J2000). The right panel 
displays the spatial distribution of members in the $10^{\prime} \times 
10^{\prime}$ region around NGC 6611. The cross and solid and dashed orange circles represent 
the center of NGC 6611, $r_h$ and $r_{cl}$, respectively.}\label{fig5}
\end{figure*}

\subsection{Radio observations}\label{sec23}
In order to compare the kinematics of stars with that of cold molecular gas, 
we observed the molecular line $^{13}$CO $J=$1$-$0 
at 110.20135~GHz toward an 1~$\degr$ $\times$ 1~$\degr$ 
region centered on M16 (R.A.: 18$^h$18$^m$48.$^s$0; decl.: 
$-$13$\degr$48$^m$24.$^s$0; J2000) using the 13.7~m radio antenna 
in Taeduk Radio Astronomy Observatory (TRAO) in Daejeon, South 
Korea \citep{JKJ19} from 2019 December 10 to December 15 (UT). The TRAO 
antenna is operating within 85 $-$ 115~GHz frequency range 
with the SEQUOIA-TRAO receiver which has 16 beams arranged in 
a 4 $\times$ 4 array \citep{EGE99}. The size of the main beam is 
about 49~$\arcsec$ at 110.201~GHz. The backend fast fourier transform 
spectrometer is available for a spectral bandwidth of 62.5~MHz with a spectral 
resolution of 15~kHz, resulting in $\sim$ 170~km~s$^{-1}$ velocity 
coverage with a velocity resolution of 0.41~km~s$^{-1}$ at 110.201~GHz. 
We tuned the antenna to set the velocity coverage centered on 
23~km~s$^{-1}$.  

Our target region was divided into nine subtiles, each 
of which covers 20~$\arcmin$ $\times$ 20~$\arcmin$ area.
Each subtile was scanned in the On The Flying (OTF) mode along 
the R.A. and decl.. The position-position-velocity cube data 
with regularly spaced grids (with a 20~$\arcsec$ cell size) 
were created via post-processing using the OTFTOOL software. 
Because the OTF mode shares the off-position data and 
calibration properties during a scan, there is a scanning 
noise feature aligned along the scanning direction in the final 
cube data. In order to reduce the scanning noise, we combined 
multiple OTF data observed along the R.A. and decl. \citep{EG88} 
using the GILDAS/CLASS software\footnote{\url{https://www.iram.fr/IRAMFR/GILDAS/}}.
We repeated this procedure until the rms noise level reaches 
about 0.5~K for the $^{13}$CO line. The total integration time 
is about 25 hours for our OTP mapping observations. 

Baseline signals were fit to the first order polynomial 
functions and then subtracted from the observed lines. 
The final position-position-velocity cube data constructed 
from the $^{13}$CO line have the rms noise 
temperature of 0.524 ~K.

\subsection{Auxiliary photometric data}\label{sec24}
$UBV$ photometry is useful to estimate reddening toward 
young stellar systems because the reddening vectors 
between $U-B$ and $B-V$ colors are well established 
\citep{SLB13}. We obtained the $UBV$ photometric data 
of 11 early-type members from the previous studies 
\citep{HJ56,HM69,HJI61}. The reddening $E(B-V)$ of 
individual early-type members can be obtained in 
the ($U-B$, $B-V$) diagram.

We also obtained the near-infrared $JHK_S$ 
photometric data from the Two Micron All Sky 
Survey (2MASS; \citealt{SCS06}). Since the 
early-type stars are bright, the qualities 
of their photometric data are sufficiently 
high in all passbands. These data are 
used in combination with the $UBV$ photometric 
data to investigate the reddening law toward 
M16.

\section{Results} \label{sec3}
\subsection{Spatial distribution of stars and gas}\label{sec31}
\subsubsection{Overall structure}\label{sec311}
Figure~\ref{fig5} displays the spatial distributions of 
the members and the gas in M16. The strong concentration 
of members is found in the center of this SFR, which corresponds 
to the young open cluster NGC 6611. In addition, 
there is a distributed stellar population spread across 
this SFR. There is no significant 
difference between the spatial distributions of 
stars with good (red dots) and poor (blue dots) 
astrometric parameters.

The gas close to the cluster is glowing brighter 
than the outskirt of M16, meaning that the gas 
is exposed to the intense ultraviolet radiation 
from massive stars in the cluster. Diffuse gas with 
low-surface brightness is extended toward the 
southern part of M16, and the northern part 
appears to be obscured by remaining molecular gas 
(contours). There are several pillars that extend 
out from the southern and northeastern edges 
of the H {\scriptsize II} bubble. The northeastern 
pillar contains molecular gas given the detection 
of the $^{13}$CO line, while the molecular line 
was not detected in the southern pillars. 

\begin{figure}[t]
\epsscale{1.0}
\plotone{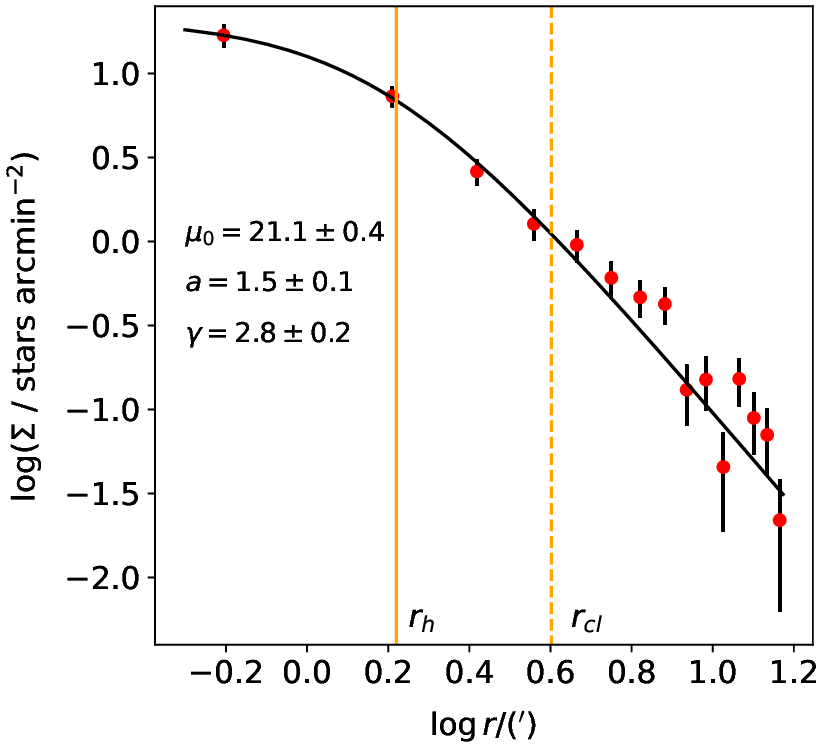}
\caption{Radial surface density profile. The Poisson 
statistics was assumed to obtain the error for 
given surface density. The solid and dashed vertical 
lines indicate the position to $r_h$ and $r_{cl}$, 
respectively. The black curve was obtained from the 
parameters ($\mu_0 = 21.1\pm0.4$, $a = 1.5\pm0.1$, 
and $\gamma = 2.8\pm0.2$) that best fit the model 
density profile \citep{EFF87}.}\label{fig6}
\end{figure}

\subsubsection{NGC 6611}\label{sec312}
In order to determine the extent of NGC 6611, we 
searched for the center of the cluster. First, the 
cluster center was visually determined in Figure~\ref{fig5} 
and then adjusted by taking the median positions 
of stars within a radius of 3$^{\prime}$ 
from the center position. The coordinate of the 
cluster center is R.A.$= 18^{\mathrm{h}} 
\ 18^{\mathrm{m}} \ 41_{\cdot}^{\mathrm{s}}49$, 
decl.$= -13^{\circ} \ 47^{\prime} \ 18\farcs1$ (J2000). 

Subsequently, we computed the radial surface density 
profile as shown in Figure~\ref{fig6}. The surface 
density decreases with respect to the projected 
distance from the cluster center. To mathematically 
characterize the structure of this cluster, the density 
profile [$\mu(r) = \mu_0 (1+r^2/a^2)^{-\gamma/2}]$ 
of \citet{EFF87} was fit to the observed one. The 
best-fit parameters are shown in Figure~\ref{fig6}. 
NGC 6611 shows a strong concentration of stars 
in its center, when compared to the surface density 
profiles of the young clusters IC 1805 and NGC 2244 
\citep{SBC17,LNH21}. The surface density of 
the inner region appears to follow the model 
profile, but the outer region ($r > 4^{\prime}$) 
tends to show overdensities. This density 
enhancement is due to the spatial distribution 
of stars extending eastward (see the right 
panel of Figure~\ref{fig5}). In this study, we 
confined the cluster to a circular region within 
a radius of $4^{\prime}$ from the center, which is 
equivalent to 2.0 pc at 1.7 kpc. In this 
apparent radius of this cluster ($r_{\mathrm{cl}}$), 
the half-number radius ($r_h$) is determined 
to be about $1\farcm7$ (0.8 pc at 1.7 kpc). 

\begin{figure}[t]
\epsscale{1.0}
\plotone{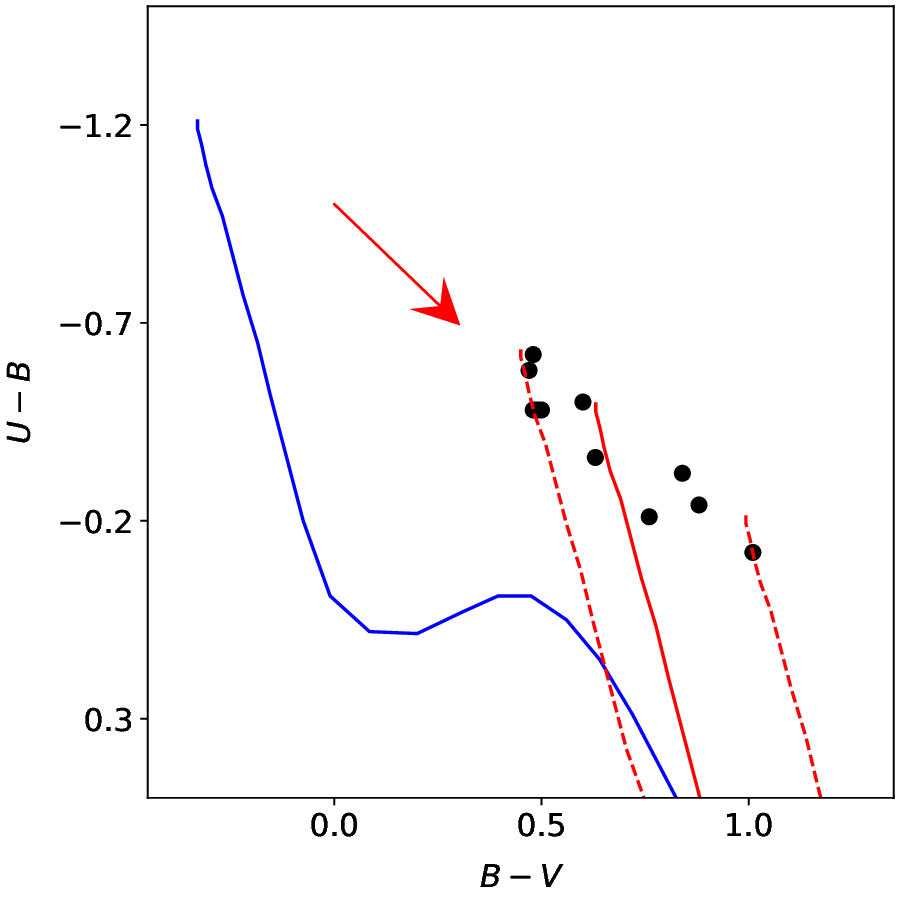}
\caption{Color-color diagram of early-type members (black dots). 
The blue solid curve represents the intrinsic color-color 
relation \citep{SLB13}, while the red solid curve shows 
the same relation reddened by the mean reddening of 0.96. 
The reddening values obtained from the early-type 
members range from 0.78 to 1.32 (dashed curves). The red 
arrow indicates the reddening vector.
respectively.}\label{fig7}
\end{figure}

\subsection{Reddening and distance}\label{sec32}
The intrinsic color relations of early-type stars and 
the reddening vector in the ($U-B$, $B-V$) diagram 
are well established \citep{SLB13}. 
Figure~\ref{fig7} displays the color-color diagram 
of the early-type members with $UBV$ photometric 
data. We determined the reddening of these early-type 
members along the reddening vector $E(U-B)/E(B-V) = 
0.72 + 0.025E(B-V)$ in the diagram. Their reddening 
ranges from 0.78 to 1.32, which quantitatively confirms 
the severe differential reddening across M16. The mean 
value is $\langle E(B-V)\rangle = 0.96 \pm 0.18$. These 
results are consistent with the mean reddening 
($\langle E(B-V)\rangle \sim 0.80$) obtained by some previous 
studies \citep{BKP99,BSB06}. 

We investigated the reddening law toward M16 
using the color excess ratios. To do this, the 
multiple colors of the early-type members were 
obtained by combining the $UBV$ photometric 
data \citep{HJ56,HM69,HJI61} and the 2MASS 
data \citep{SCS06}. The intrinsic colors 
($V-\lambda$, where $\lambda$ are $J$, 
$H$, and $K_S$) of these stars were obtained 
by interpolating their intrinsic color $(B-V)_0$ 
to the intrinsic color ($V-\lambda$) relations 
with respect to $(B-V)_0$ \citep{SLB13}. We 
then calculated the color excess $E(V-\lambda)$ 
from the observed colors minus the intrinsic 
colors.

\begin{figure*}[t]
\epsscale{1.0}
\plotone{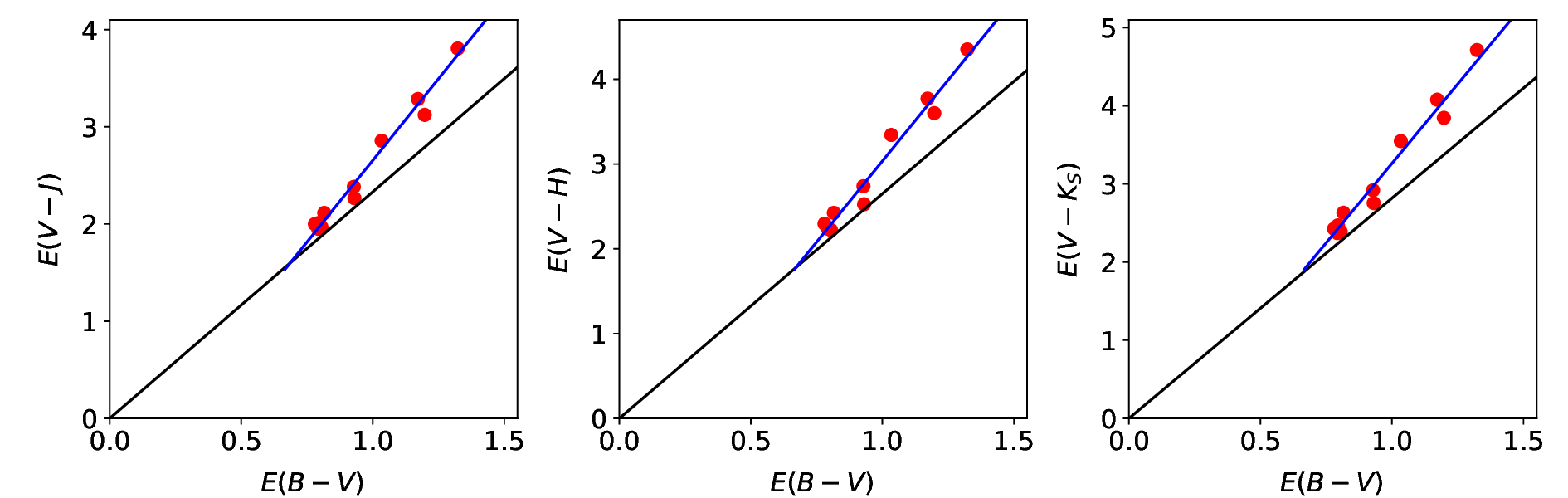}
\caption{Color excess ratios of the early-type 
members (red dots). The black line shows the 
color excess ratios following the normal 
reddening law ($R_V = 3.1$), while the blue line 
corresponds to $R_V = 4.5$, which represents 
the abnormal reddening law.}\label{fig8}
\end{figure*}

Figure~\ref{fig8} displays the color excess 
ratios obtained from the early-type members. 
The color excess ratios of these stars do not 
follow the normal reddening law corresponding 
to $R_V = 3.1$ (black line). The color excess 
$E(V-\lambda)$ at given $E(B-V)$ are greater 
than those expected from the normal reddening 
law. We calculated the $R_V$ for the intracluster 
medium (hereafter $R_{V,cl}$) using the equations 
6--8 of \citet{SLB13}, which are the same as the 
relations presented in \citet{GV89}. The mean 
$R_{V,cl}$ is $4.5\pm0.1$ (blue line), 
indicating the abnormal reddening law for the 
intracluster medium. This result supports the 
previous estimates in an $R_{V,cl}$ range from 
3.2 to 4.8 \citep{CW90,TWF90,HMS93}. 

While the reddening law in the intracluster medium 
appears to be abnormal, the foreground medium likely 
follows the normal reddening law \citep{HMS93}. 
Assuming the normal reddening law for the 
foreground medium, we estimated the maximum 
foreground reddening $E(B-V)_{fg}$ (or the 
minimum reddening of the intracluster medium) 
from the intersection of the two straight 
lines corresponding to an $R_{V,cl}$ of 
4.5 and the foreground $R_V$ of 3.1 in 
Figure~\ref{fig8}. The mean value of the maximum 
foreground reddening calculated from the three 
different color excess ratios is $\langle E(B-V)_{fg} 
\rangle = 0.67 \pm 0.02$ (s.d.).

\begin{figure}[t]
\epsscale{1.0}
\plotone{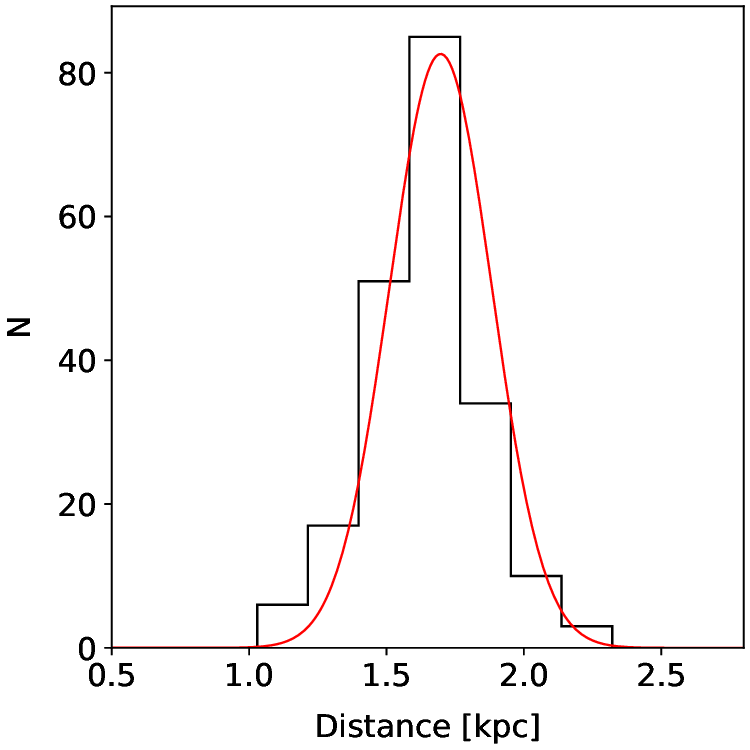}
\caption{Distance distribution of member stars (black). Stars 
with good astrometric parameters (RUWE $<$ 1.4 
and duplicate flag $=$ 0) were used to compute 
distances. The red curve represents the Gaussian 
fit to the distance distribution. The distance to 
M16 is determined to be $1.7 \pm 0.2 $ (s.d.) 
kpc from the Gaussian center.}\label{fig9}
\end{figure}

The distances of the individual members were 
obtained from the inversion of their parallaxes. 
The stars with poor astrometric parameters were 
not used for the distance calculation. Figure~\ref{fig9} 
exhibits the distribution of the distances. We 
fit a Gaussian to the distance distribution and 
determined the distance of $1.7 \pm 0.2$ kpc 
from the center of the best-fit Gaussian (red 
curve). This result is consistent with the 
distances derived from the Gaia parallaxes 
by recent studies \citep{CJV18,KHS19}.

\begin{figure*}[t]
\epsscale{1.0}
\plotone{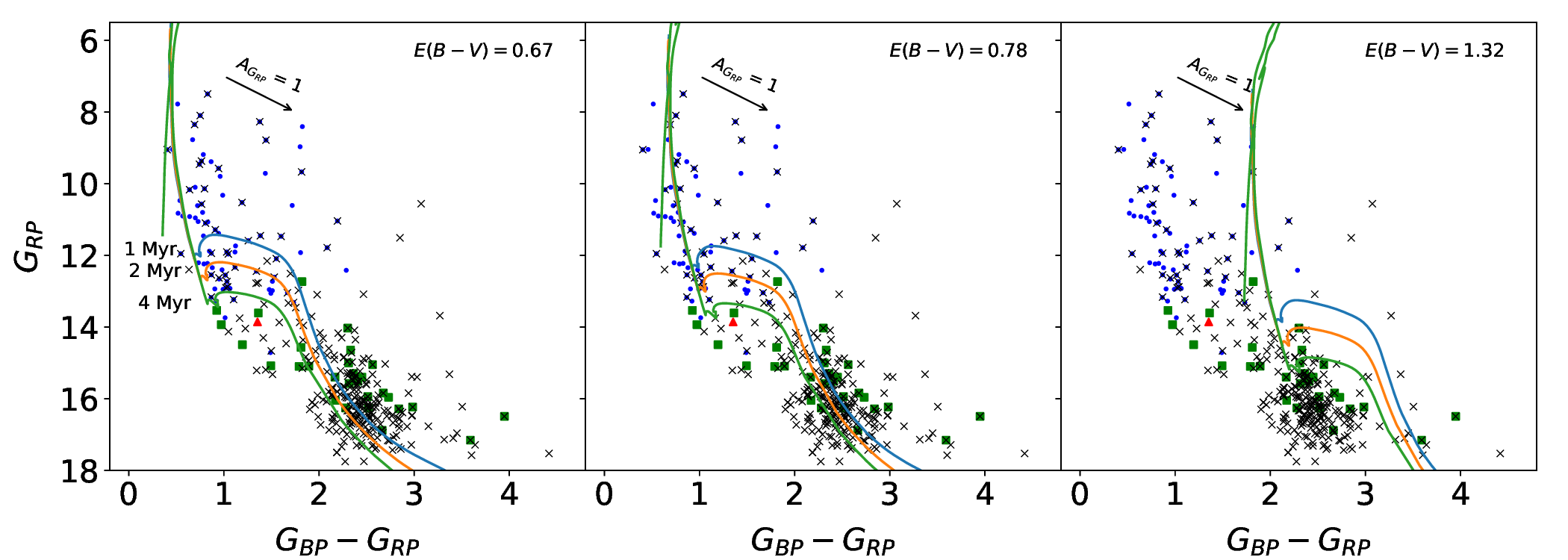}
\caption{CMDs in comparison with 
the MESA theoretical isochrones for three different ages 
1 Myr (blue), 2 Myr (orange), and 4 Myr (green) 
\citep{CDC16,D16}. The isochrones are reddened by 
$E(B-V) = 0.67$, 0.78, and 1.32 from the left panel 
to right panel, respectively. The arrow indicates the reddening 
vector corresponding to $A_V = 1$ mag. }\label{fig10}
\end{figure*}

\subsection{Age}\label{sec33}
We estimated an age range of stars distributed 
in M16 using the MESA theoretical isochrones 
\citep{CDC16,D16}. The total extinction in the 
$G_{RP}$ band and the reddening $E(G_{BP} - G_{RP})$ 
were obtained from the empirical relations obtained 
by Lee et al. (in prep.) as 
below;

\begin{equation}
 \begin{aligned}
 R_V &= 11.1 E(G_{BP}-V)/E(B-V) \\
      &= 2.65 E(V-G_{RP})/E(B-V).
 \end{aligned}
\end{equation}

\noindent The foreground medium toward M16 
follows the normal reddening law ($R_V = 3.1$), while 
the reddening of the intracluster medium is explained 
by an abnormal law ($R_{V,cl} = 4.5$). Therefore, the 
total extinction ($A_{G_{RP}}$) is, then, the sum of the 
foreground extinction ($A_{G_{RP},fg}$) and the intracluster 
extinction ($A_{G_{RP},cl}$) as follow; 

\begin{equation}
\begin{aligned}
A_{G_{RP}} &= A_{G_{RP},fg} + A_{G_{RP},cl} \\
        &= 0.623R_{V,fg}E(B-V)_{fg} \\
        &+ 0.623R_{V,cl}[E(B-V) - E(B-V)_{fg}].
\end{aligned}
\end{equation}

Figure~\ref{fig10} displays the CMDs of members, with 
the MESA isochrones for 1, 2, and 4 Myr superimposed on 
the CMDs. We applied three different reddening 
values to each set of isochrones: 0.67, 0.78, and 1.32 
(from left-hand panel to right-hand panel). As a result, 
the isochrones with the reddening of $E(B-V) = 0.78$ 
fit the overall distribution of members well (middle 
panel), while the smallest and largest reddening values 
match the blue and red ridges of the CMD, respectively 
(left and right-hand panels). In all cases, the majority 
of members are found between the isochrones for 1 Myr 
and 4 Myr. This age range is consistent with those of 
previous studies \citep{HMS93,BSB06,GPM07,SKK23}.

Some members appear to be older or younger than most stars 
when compared to the isochrones, possibly indicating an age 
spread among stars in M16. The existence of age spread in SFRs 
has long been debated; if the age spread does indeed exist, the 
age range of stars is related to the formation timescale of stellar 
systems \citep{PS99,PS00,PS02}. Some stars may actually 
be younger or older than the main population of M16; however, a 
number of factors, variability, multiplicity, differential 
reddening, etc. can lead to over- or underestimation of the 
ages of stars \citep{H99,H01,H03}. In addition, some pre-main 
sequence stars may be obscured by their edge-on disks, 
causing them to fade and overestimate their age \citep{SB10}. 

\begin{figure}[t]
\epsscale{1.0}
\plotone{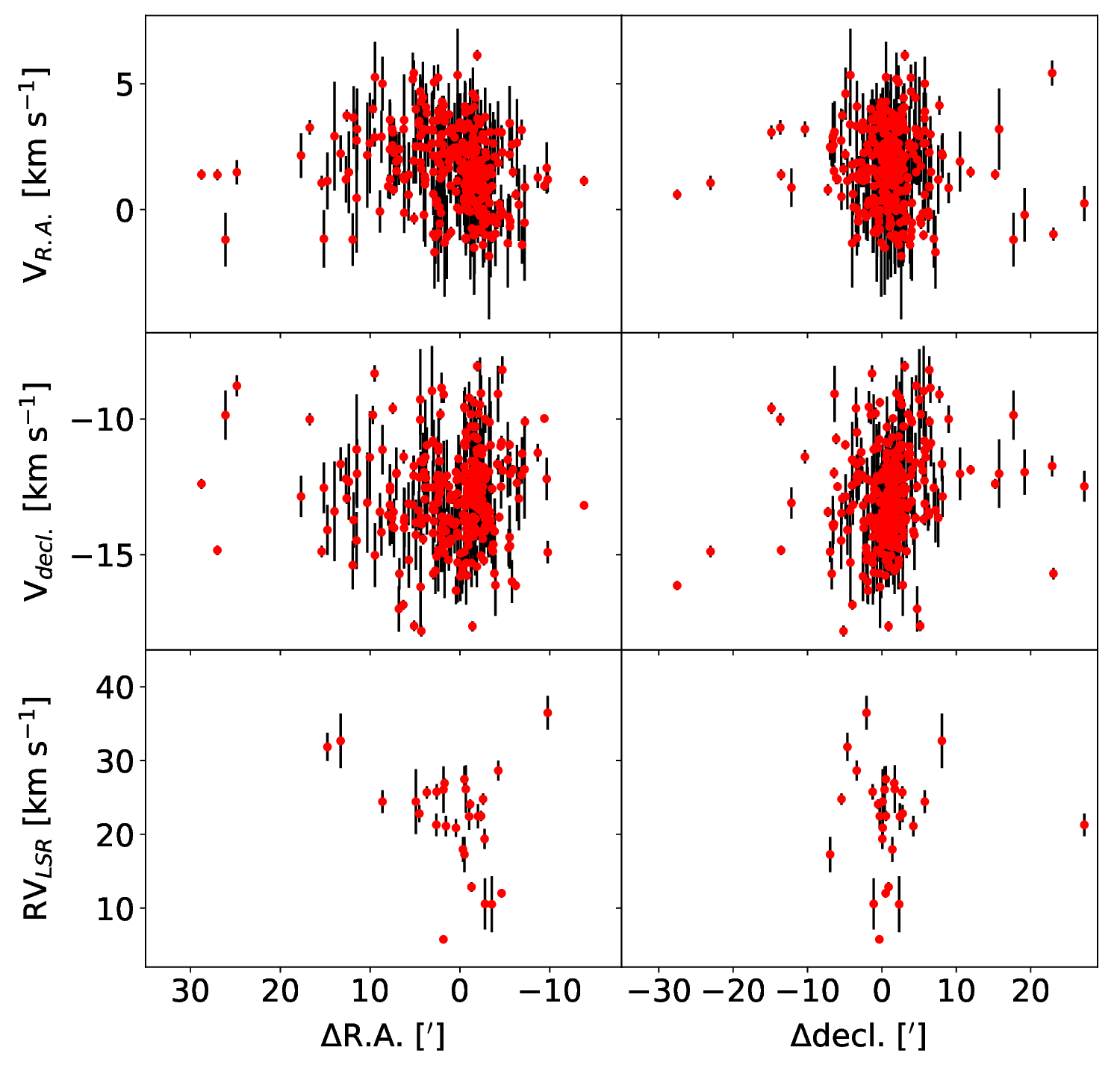}
\caption{Distributions of tangential velocities 
($V_{\mathrm{R.A.}}$ and $V_{\mathrm{decl.}}$) and RVs 
with respect to R.A. and decl. in M16. The vertical lines 
represents the errors of velocities. }\label{fig11}
\end{figure}

\subsection{Kinematics}\label{sec34}
We investigated the kinematic properties of the members 
in M16 using PM and RV data. The PMs of members can be 
affected by projection effects caused by their motions 
along the line of sight. This depends on the distance 
and size of their host cluster \citep{vL09}. We corrected 
for the projection effects on the PMs of members using 
the equation 13 of \citet{vL09}. In addition, 
there is a correction term for the RVs of stars based on their 
positions relative to the cluster center and the systemic PM. 
The correction values are, on average, 0.5 km s$^{-1}$, 
which is smaller than the RV errors in the measurement of this study. Therefore, we did not apply the correction values 
to the observed RVs.

We computed the tangential velocities ($V_{\mathrm{R.A.}}$ 
and $V_{\mathrm{decl.}}$) of stars along R.A. and decl. by 
multiplying the PMs by the distance of 1.7 kpc. Figure~\ref{fig11} 
displays the distributions of the velocities with respect to R.A. 
and decl.. $V_{\mathrm{R.A.}}$ and $V_{\mathrm{decl.}}$ range 
from $-2$ km s$^{-1}$ to 5 km s$^{-1}$ 
[$\langle V_{\mathrm{R.A.}} \rangle = 1.7 \pm 1.6$ (s.d.) km s$^{-1}$] 
and from $-17$ km s$^{-1}$ to $-8$ km s$^{-1}$ [$\langle V_{\mathrm{decl.}}\rangle = -12.9 \pm 1.7$ (s.d.) km s$^{-1}$], respectively. The members observed 
in this study have RVs ranging from 10 km s$^{-1}$ to 35 km s$^{-1}$ [$\langle 
\mathrm{RV}_{\mathrm{LSR}} \rangle = 22.4 \pm 6.6$ (s.d.) 
km s$^{-1}$]. There is no detectable systematic 
variation across M16 compared to the velocity spread.

\subsubsection{Expansion of NGC 6611}\label{sec35}
To better determine the systemic motion of NGC 6611, we 
computed median PMs of members within $r_h$. The median 
values are $0.200$ mas yr$^{-1}$ in R.A. and $-1.618$ 
mas yr$^{-1}$ in decl., corresponding to 
$V_{\mathrm{R.A.}} = 1.6$ km s$^{-1}$ and $V_{\mathrm{decl.}} 
= -13.0$ km s$^{-1}$, respectively. These values are 
consistent with those of \citet{CJV18} 
and have been adopted as the systemic PMs of the 
central cluster NGC 6611. The median RV of this cluster 
is about 22.4 km s$^{-1}$.
  
\begin{figure}[t]
\includegraphics[width=0.45\textwidth]{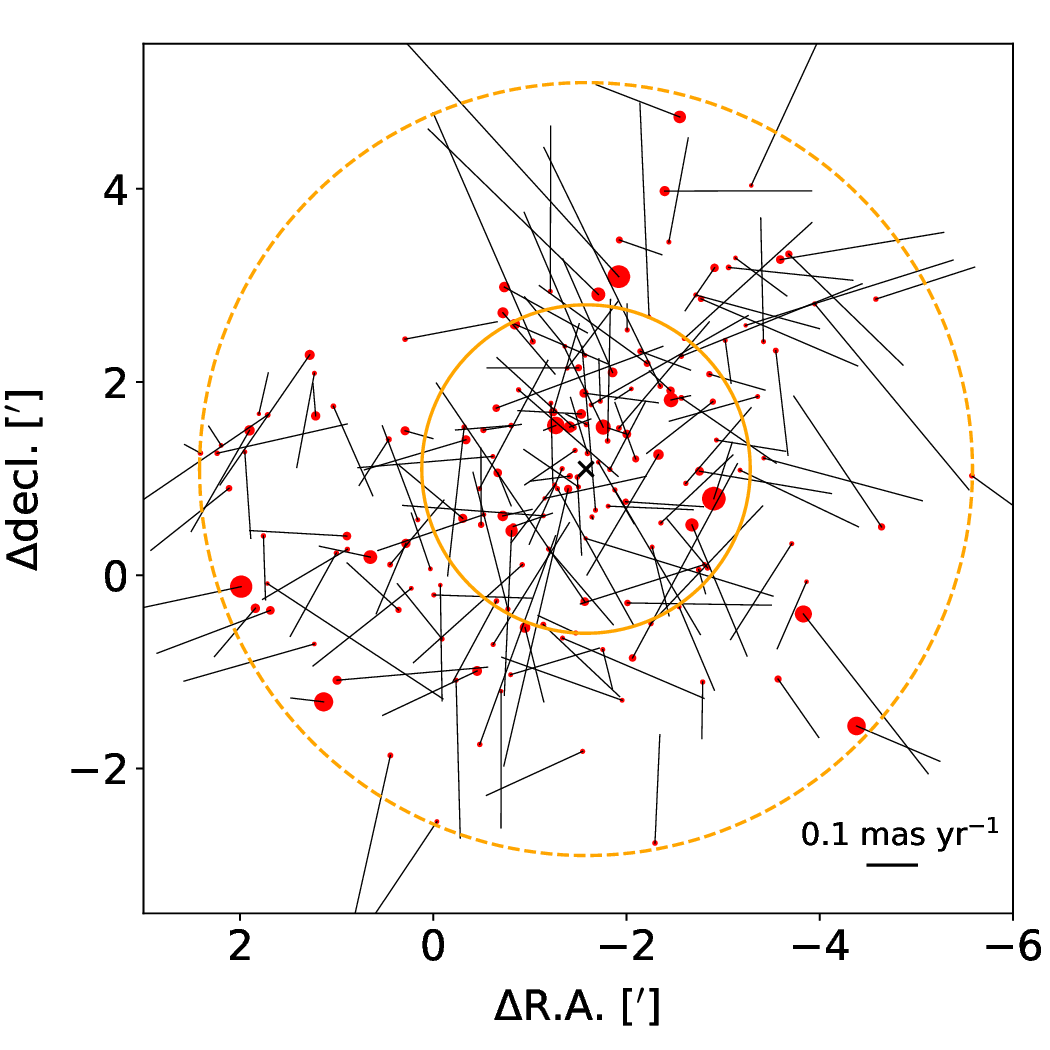}
\includegraphics[width=0.47\textwidth]{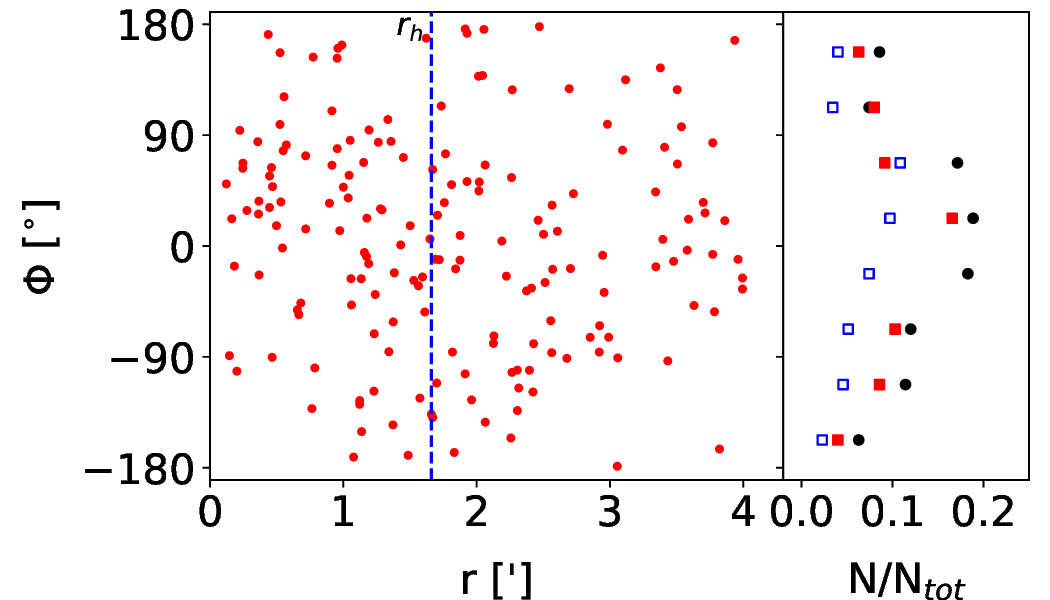}
\caption{Map of proper motion vectors (upper) 
and the $\Phi$ distribution of the cluster members 
(lower). In the upper panel, the inner (solid) and 
outer (dashed) circles represent $r_h$ and $r_{cl}$, 
respectively. The dashed line in the lower-left hand 
panel corresponds to $r_h$. The black dots, blue open 
squares and red filled squares in the lower right-hand panel 
show the histograms of $\Phi$ values for all cluster 
members, members within $r_h$, and members beyond $r_h$, 
respectively. }\label{fig12}
\end{figure}

The upper panel of Figure~\ref{fig12} displays the PMs 
of members relative to the systemic motion of NGC 6611. 
The members within $r_h$ (inner circle) show no clear 
pattern of motions, while the other members beyond the 
radius tend to move away from the center of the cluster. 
We calculated the vectorial angles ($\Phi$) of individual 
members to quantitatively analyze their PMs relative to 
the central cluster as done in our previous studies 
\citep{LNG19,LHY20,LNH21,LNH22,LHL23}. The $\Phi$ is 
defined by the angle between the position vector 
from the center of the cluster and the relative 
PM vector of a given star. A $\Phi$ of 180$^{\circ}$ 
or $-180^{\circ}$ indicates that a star is sinking 
toward the cluster center, while a $\Phi$ of 0$^{\circ}$ 
means that it is raidally escaping outward.

The lower-left panel of Figure~\ref{fig12} exhibits 
the $\Phi$ distribution of members with respect 
to the projected distance from the center of the 
cluster. A number of stars have $\Phi$ 
values between $-90^{\circ}$ and $90^{\circ}$. 
In the lower-right panel, we examined 
the $\Phi$ distributions for all stars (black dots) 
and for stars in and out of $r_h$ (blue open squares and 
red filled squares), respectively. These histograms have peaks 
in $\Phi$ around $0^{\circ}$, indicating an 
expansion of the cluster. This pattern appears 
more pronounced beyond $r_h$.

\begin{figure}[t]
\includegraphics[width=0.47\textwidth]{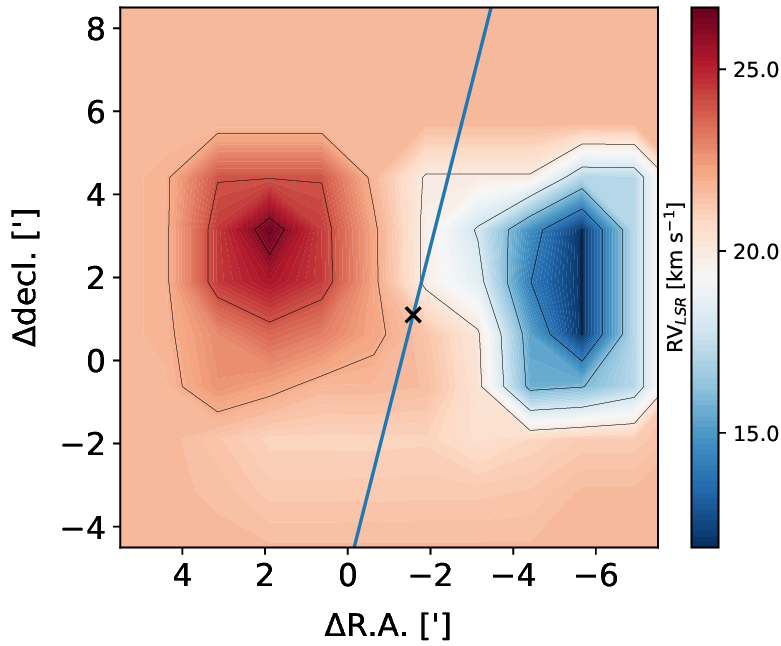}
\includegraphics[width=0.47\textwidth]{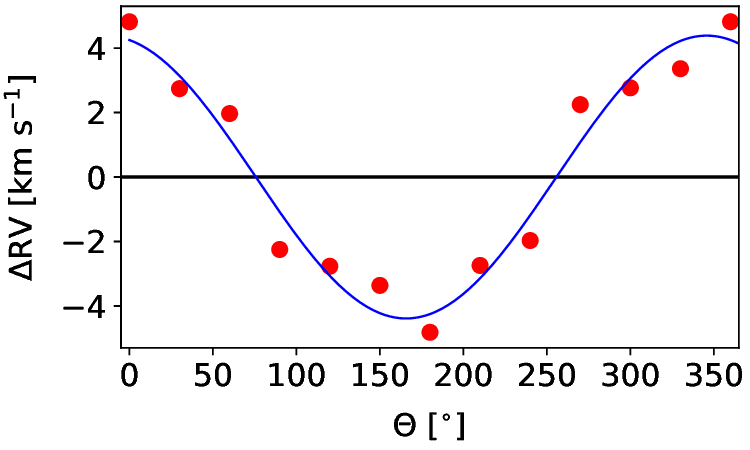}
\caption{Rotation of NGC 6611 within $r_{cl}$. The 
upper panel represents the spatial variation of mean 
RVs that were computed within spatial bins of 7\farcm0. 
The cross and solid line indicate the 
center of NGC 6611 and the orientation of the projected 
rotational axis. The reference coordinate is the same as 
Figure~\ref{fig5}. In the lower panel, the dots represents the 
differences of the mean RVs between two regions divided 
by arbitrary axes passing through the center of the 
cluster. $\Theta$ is the position angles of the arbitrary 
axes. The blue curve shows the best-fit sinusoidal 
curve.}\label{fig13}
\end{figure}

\subsubsection{Cluster rotation}\label{sec36}
The signature of cluster rotation can be 
investigated using the spatial distribution of RVs. 
The upper panel of Figure~\ref{fig13} shows the smoothed 
spatial distribution of mean RVs. A clear pattern 
of rotation is detected. To measure the rotational 
velocity of this cluster, we applied the same method 
as used in previous 
studies \citep{LKL09,MDFY13,LNH21} to our data. An 
axis passing through the center of the 
cluster was considered, and its initial position angle ($\Theta$) 
was set to $0^{\circ}$--$180^{\circ}$ (north-south). We 
computed the difference between the mean RVs of stars 
in the two regions separated by the axis. Then, the same 
procedure was repeated for stars within a radius of $4\farcm0$ 
by changing the position angles of the axis in a range 
from $0^{\circ}$ to $360^{\circ}$ with an interval 
of $30^{\circ}$. If the cluster is rotating, the 
differences of mean RVs appears to systematically vary 
with respect to the position angles of the axis. The 
maximum or minimum difference occurs when the axis 
is aligned with the rotational axis projected on the sky.

The lower panel of Figure~\ref{fig13} shows the 
systematic variation of the mean differences of RVs. 
This variation was fit to the sinusoidal curve as below:
\begin{equation}
\Delta \langle\mathrm{RV}\rangle = A \sin(\Theta + \Theta_0) 
\end{equation}
\noindent where $A$ and $\Theta_0$ represent the amplitude
and phase, respectively. Half of the amplitude corresponds 
to the projected rotational velocity ($V_{\mathrm{rot}}\sin i$). 
If the rotational axis is perpendicular to the line of sight 
$i = 90^{\circ}$, the rotational velocity $(V_{\mathrm{rot}})$ 
of this cluster is $2.2 \pm 0.2$ km s$^{-1}$. Therefore, this 
value is the lower limit of $(V_{\mathrm{rot}})$. 

The orientation of the projected rotational axis in the sky 
was inferred from $270^{\circ} - \Theta_0$. The position angle 
of the projected rotational axis is $166\pm4^{\circ}$ 
(from north to east), which means that the rotational 
axis lies in the northwest-southeast direction.

\begin{figure*}[t]
\epsscale{1.0}
\plottwo{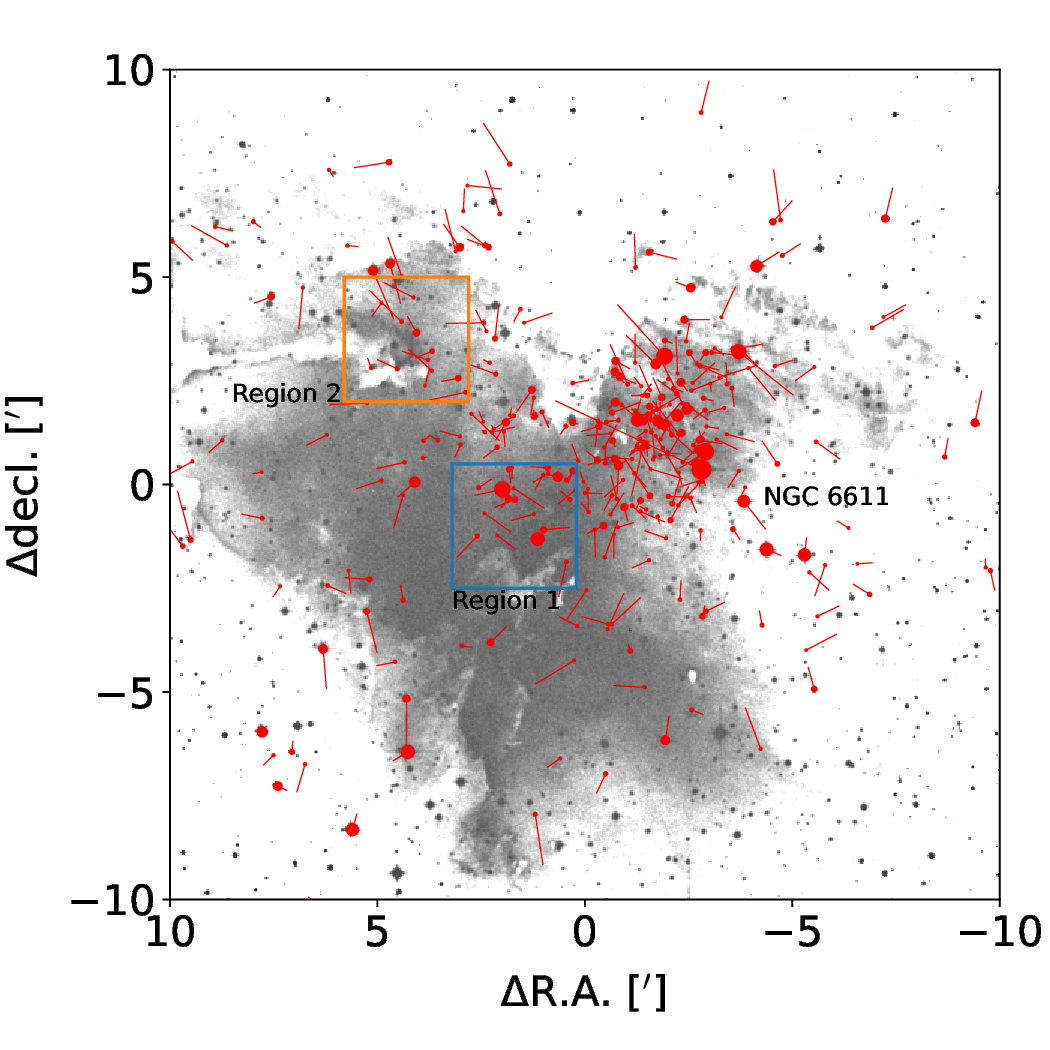}{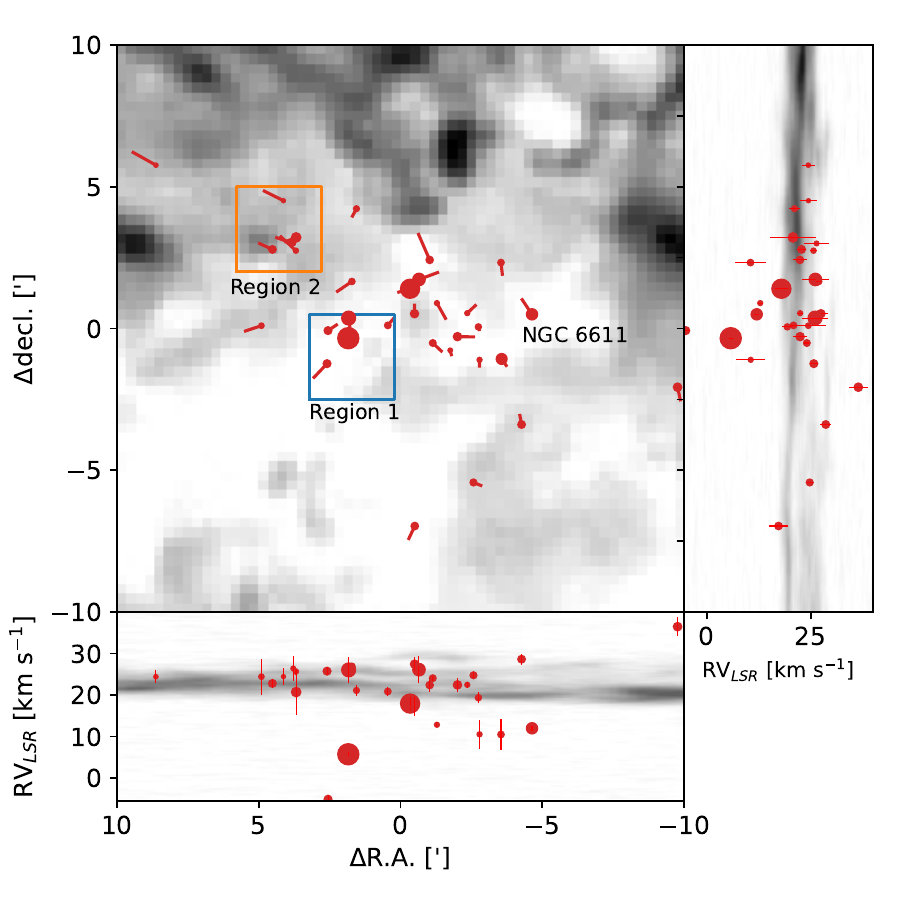}
\caption{Kinematics of stars in M16. The left-hand panel 
displays the proper motion vectors of stars relative to 
NGC 6611. The gray scale in the right-hand panel exhibits 
the integrated intensity map and position-velocity 
diagrams (gray scales) of the $^{13}$CO $J = 1 - 0$ line. 
The stars with RV measurements are 
superimposed on the figure. The vertical and horizontal 
lines in the position-velocity diagrams represent the errors of 
stellar RVs. The orange and blue boxes 
contain two prominent gas pillars that are likely to be 
physically associated with neighboring stars.}\label{fig14}
\end{figure*}

\subsubsection{A distributed population}\label{sec37}
The left-hand panel of Figure~\ref{fig14} displays the 
PMs of members relative to the systemic motion of NGC 6611. 
The distributed stellar population does not show systematic 
outward motions from the cluster and stream 
motions on the scale of tens of parsecs. Their PMs appear 
to be randomly oriented. The right-hand panel compares the RVs of 
stars with those of molecular gas traced by the 
$^{13}$CO $J = 1 - 0$ transition line. The RVs of the 
molecular gas range from 20 km s$^{-1}$ to 
30 km s$^{-1}$. Most members follow the kinematics 
of the remaining gas.

\begin{figure*}[t]
\epsscale{1.0}
\plottwo{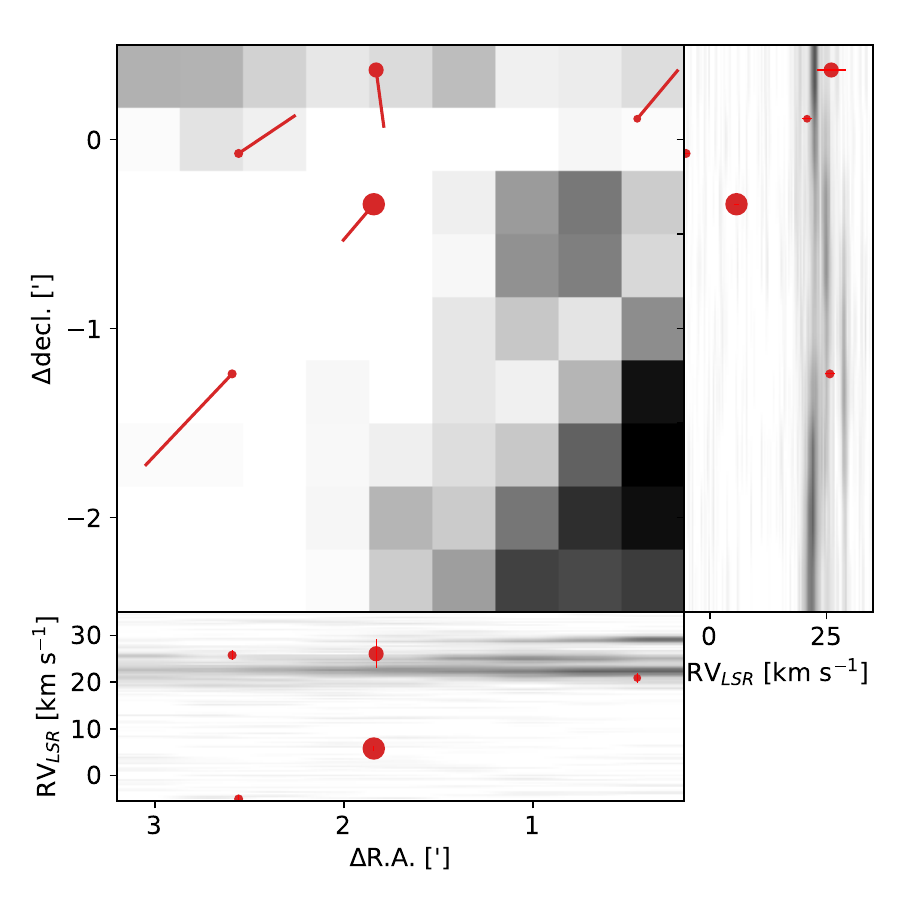}{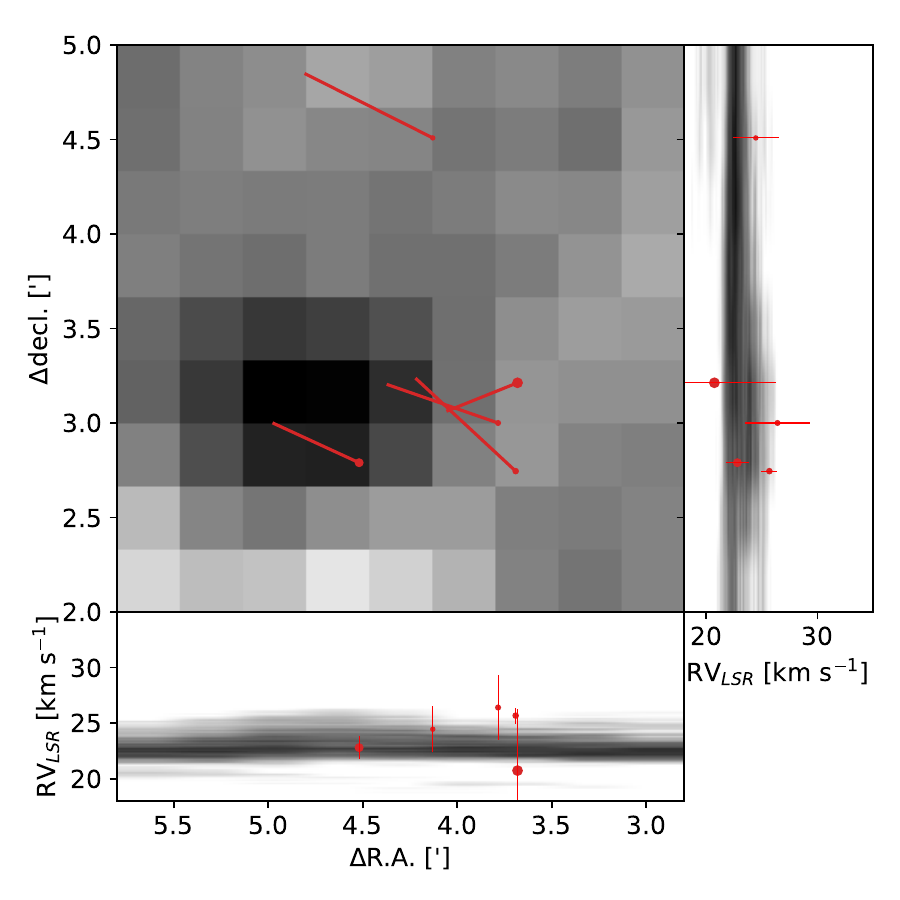}
\caption{Integrated intensity maps and position-velocity 
diagrams of stars and gas in the Regions 1 (left) and 2 
(right). The symbols are the same as Figure~\ref{fig14}.}\label{fig15}
\end{figure*}

The spatial distribution of members appears to extend 
from the cluster to the southern pillar (Region 1). In 
addition, there is a small group of members in the vicinity 
of the northeastern pillar (Region 2). We investigated 
the physical relationship between the gas structure and the members 
in the two regions. The PM vectors of 
the members in the Region 1 show no systematic motion 
in the left-hand panel of Figure~\ref{fig15}. 
We then compared their RVs with those of the southern 
pillars. However, these gas structures are invisible 
in the integrated intensity map of the $^{13}$CO 
$J = 1 - 0$ line. It is impossible to compare the RVs 
of the CO gas with those of stars in the Region 1. 

\begin{figure*}[t]
\includegraphics[width=0.47\textwidth]{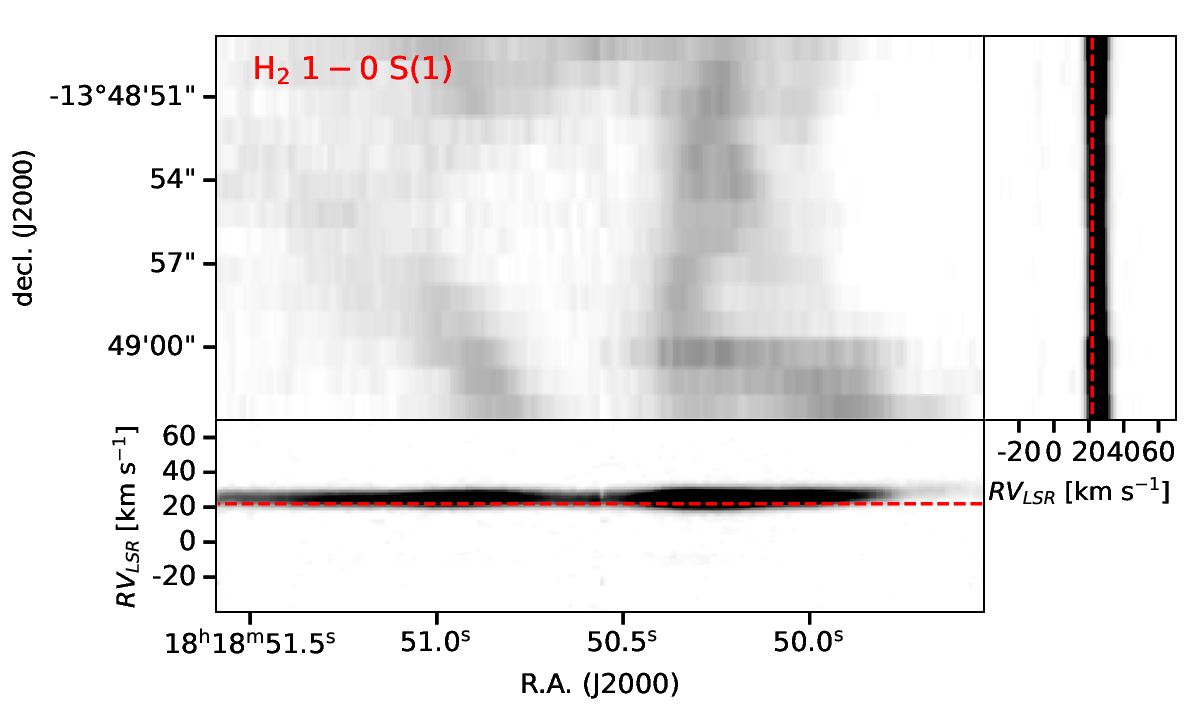}\includegraphics[width=0.47\textwidth]{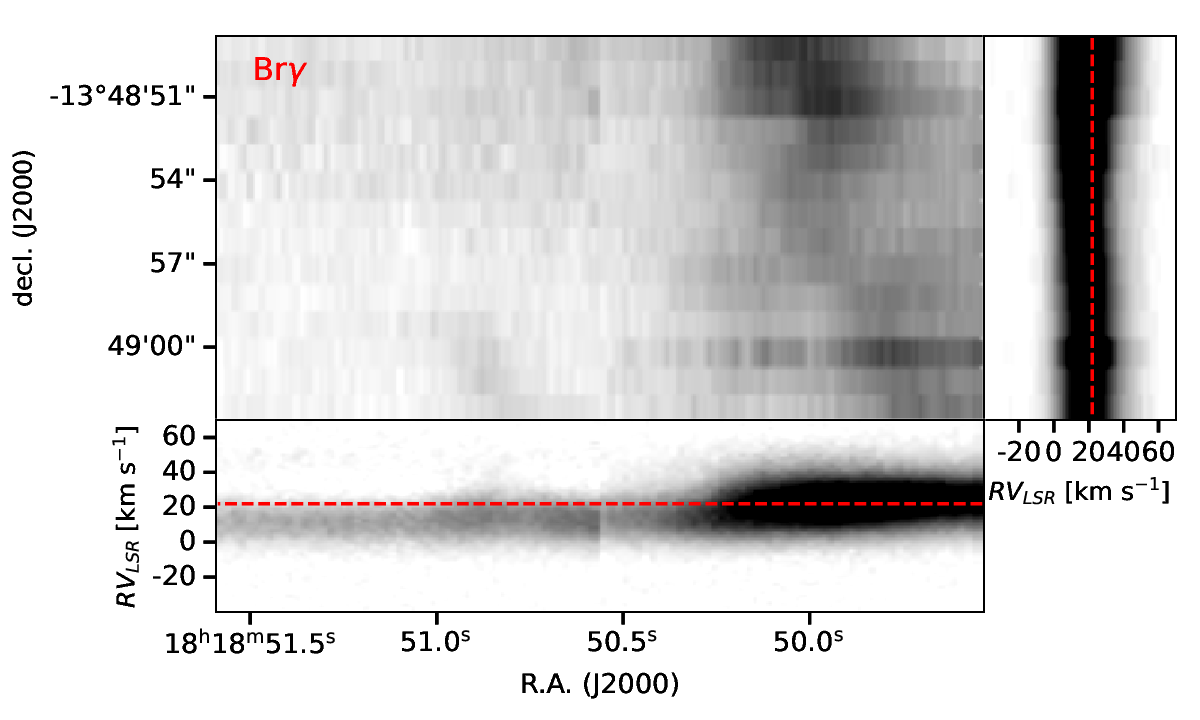} 
\includegraphics[width=0.47\textwidth]{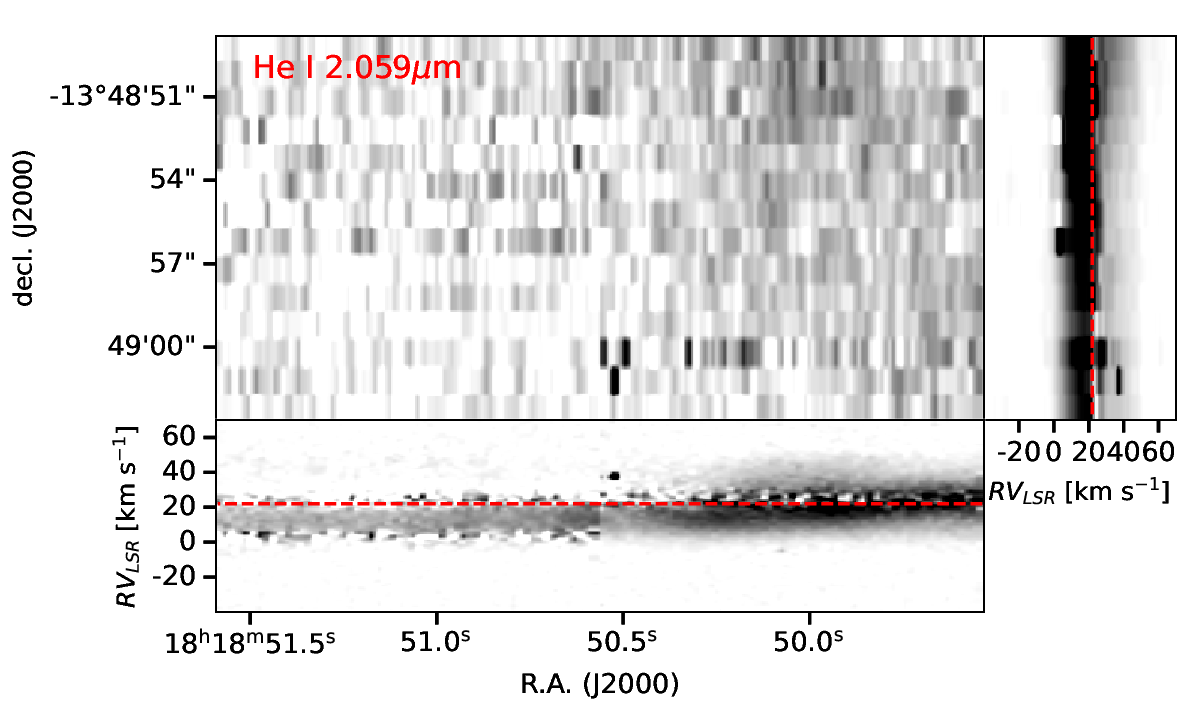}\includegraphics[width=0.47\textwidth]{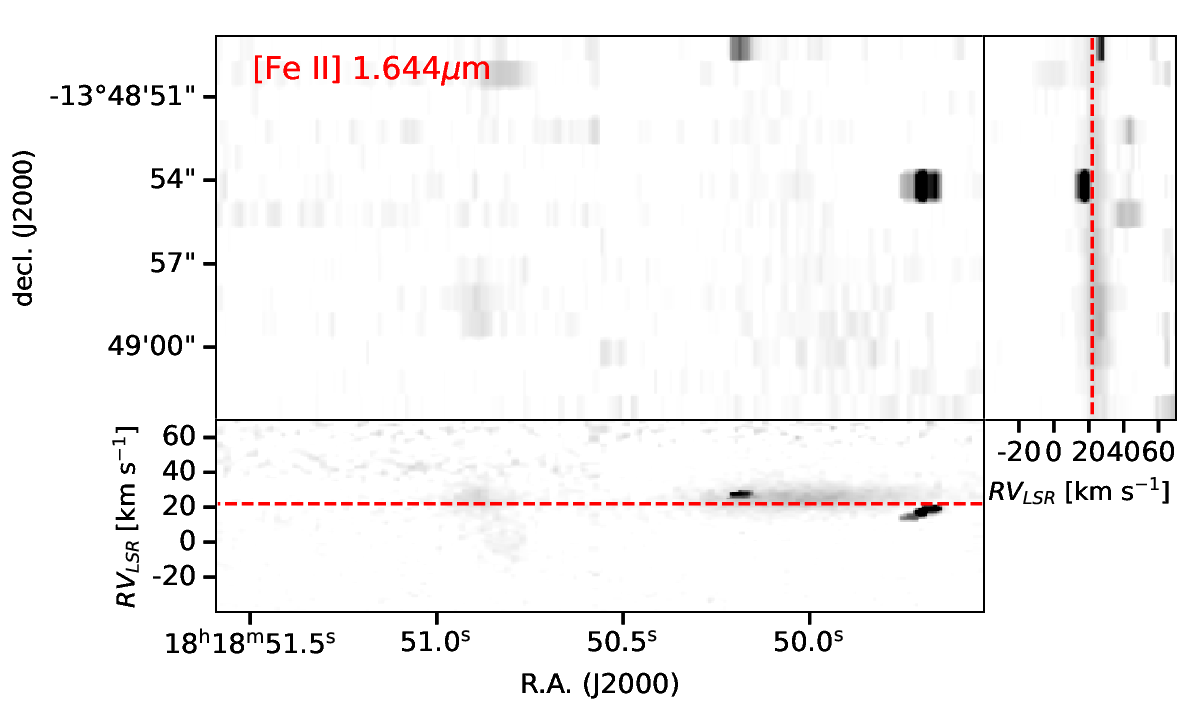}
\caption{Integrated intensity maps and position-velocity diagrams 
of the H$_2$ 1 -- 0 S(1), Br$\gamma$, He {\scriptsize I} 2.059 \micron, 
and [Fe {\scriptsize II}] 1.644 \micron \ lines. A total of 
28 slits cover the head of one southern pillar (see Figure~\ref{fig4}). 
The red dashed lines represent the system RV of NGC 6611 for comparison.}\label{fig16}
\end{figure*}

We investigated the kinematics of one of the southern 
pillars using its near-infrared spectra (see Figure~\ref{fig4}) 
from which the H$_2$ 1 -- 0 S(1), Br$\gamma$, He {\scriptsize I} 2.059 \micron, and [Fe {\scriptsize II}] 1.644 \micron\ lines were detected. 
Figure~\ref{fig16} displays the integrated 
intensity maps and position-velocity diagrams obtained 
from the near-infrared spectral cube data. The detected 
lines are commonly stronger and broader on the west side 
of the pillar head. This fact indicates the physical 
interaction between the gas and ionizing photons. 

Molecular hydrogen have almost constant RVs of about 25 
km s$^{-1}$ (see also \citealt{KPT23} for comparison), 
while hot gas traced by the atomic lines of hydrogen 
and helium exhibits a velocity gradient in the east-west direction. On the 
east side of the pillar head, these two atomic lines 
appears blueshifted by about 10 km s$^{-1}$ compared 
to the molecular hydrogen line. The hot gas likely 
surrounds the molecular hydrogen gas and is being 
blown away by radiative feedback from massive stars. 
There are three stars that are probably associated 
with the southern pillars. These stars have RVs ranging 
from 20 km s$^{-1}$ to 30 km s$^{-1}$ (the left-hand 
panel of Figure~\ref{fig15}). 

On the other hand, the members in the Region 2 are 
heading toward northeast direction, which is the opposite 
direction from NGC 6611 (the right-hand panel 
of Figure~\ref{fig15}). Therefore, they appear to be 
moving away from the cluster. The northeastern pillar 
still contains a detectable amount of CO gas in the 
integrated intensity map. It seems to be spatially 
associated with these stars. The position-velocity 
diagrams in the right-hand panel of Figure~\ref{fig15} 
compare the RV of the gas pillar with those of the stars 
in the Region 2. The gas pillar have RVs in a range 
from 23 km s$^{-1}$ 
to 26 km s$^{-1}$. Although the one star has a large 
RV error, the RVs of the stars are similar to the RVs 
of the gas pillar within the RV errors. Therefore, these 
stars are physically associated with the gas pillar.

\section{Star formation in M16}\label{sec4}
\subsection{Formation of NGC 6611}\label{sec41}
Given the ages of members, the majority of stars 
were formed 2 Myr ago. A total of 189 
members were found within $r_{cl}$ 
(Section~\ref{sec312}). If all of these members belong 
to NGC 6611, approximately 55\% of stars 
might have formed in this cluster. Similarly, most 
stars found in the Galactic OB associations tend 
to be clustered \citep[etc]{B64,KAG08,KFG14,L24}. 
Therefore, the local star formation efficiencies 
in the past regions that formed stellar clusters 
were likely higher than in other regions 
of the same molecular clouds \citep{BSC11,K12}. 

The rotation of NGC 6611 might have been 
inherited from its natal cloud. There are 
two possible scenarios. One is the monolithic 
collapse of a rotating cloud. Indeed, a 
number of molecular clouds in external 
galaxies are rotating \citep{REP03,T11,BHR20}. 
According to \citet{JKW24}, the spin rates 
of rotating open clusters are much larger 
than those expected from the {\tt N-body} 
simulations of model clusters without 
the initial rotation, indicating 
that the natal clouds of the Galactic 
open clusters rotated. Interestingly, 
however, the rotational axes of the clusters 
do not always coincide with their orbital 
angular momentum vectors. 

The other scenario explaining the formation 
of rotating clusters is the hierarchical 
merging of gaseous and stellar clumps 
in a cloud \citep{M17}. An observational 
example is the core of 30 Doradus in the 
Large Magellanic Cloud. The core has an 
elongated structure composed of the 
young massive cluster R136 and a sparse 
stellar group. There is an age difference 
between the two stellar groups. It is 
believed that these clusters are 
undergoing a merger event \citep{SLG12}. 
The rotation of R136 has been detected 
\citep{HGE12}, which is likely due to 
the merger event. 

Another example is the rotating open 
cluster NGC 2244 in the Monoceros OB2 
association \citep{LNH21}. This 
cluster has a neighboring stellar group 
to its west. These two stellar systems 
likely formed in the same natal cloud. 
If the neighboring group is rotating 
around the cluster, there is no coincidence 
between the angular momentum vectors 
of its orbit and the cluster rotation. 
The monolithic collapse may not be 
the explanation for the formation of 
NGC 2244 and its neighboring group 
of stars. 

The distributed stellar population in 
M16 do not show detectable RV variation 
across this region. The fraction of 
the distributed population is about 45\% 
of the total members. If we assume that 
its total stellar mass is similar to 
the mass of the cluster, the simple 
calculation based on the angular momentum 
conservation yields a rotational 
velocity of 0.5 km s$^{-1}$ which is 
smaller than the rotational speed 
of NGC 6611. The precision of the current 
RV data is insufficient to detect such a 
small variation. Therefore, the theoretical 
models of both monolithic collapse and 
hierarchical merger are still possible 
explanations for the formation of NGC 6611.

\subsection{Origin of the distributed population}\label{sec42}
The young open cluster IC 1805 in the SFR W4 
is undergoing expansion \citep{LHY20}. The pattern 
of expansion appears to be pronounced over $r_h$. 
The stars escaping from the cluster form a distributed 
population that is spread out over 20 pc. This 
observational fact can be explained if the cluster 
members are rebounding after cold collapse into the 
center of the cluster \citep{LHY20}.

We investigated the contribution of the expansion 
of NGC 6611 to the formation of 
distributed population. To do this, stars escaping 
from the cluster ($-90 < \Phi [^{\circ}] < 90$) 
beyond $r_h$ were considered. The mean total PM of 
these stars relative to the cluster is about 
0.260 mas yr$^{-1}$, which is equivalent to 2.1 km s$^{-1}$ 
at 1.7 kpc. Most of the members are distributed 
within a radius of 10$^{\prime}$ ($\sim$5 pc). 
The travel time of a star from the center of the cluster 
to the radius with the tangential velocity of 2.1 
km s$^{-1}$ is about 2 Myr, which is comparable to 
the age of the stars in this SFR. The expansion of 
NGC 6611 is probably associated with the formation 
of the distributed population.

Unlike IC 1805 in W4, the PM vectors of stars 
beyond $r_{cl}$ are randomly oriented, and therefore 
it cannot be concluded that the expansion of 
the cluster is the origin of the 
distributed population. Instead, the spontaneous star 
formation across M16 could be a natural explanation 
for their origin. On the other hand, star formation 
can be induced by feedback from massive stars. The 
Regions 1 and 2 are considered the 
sites of feedback-driven star formation. In 
particular, the southern pillars have steadily 
been studied at various wavelengths \citep{STN02,
FHS02,TSH02,LJM07}. The age sequence of YSOs 
and cores as well as the morphology of the pillars 
have thought to be the typical signposts 
of feedback-driven star formation. 

According to our observational results, three out 
of five stars with RV measurements seem to be 
physically associated with the pillar head. However, 
they have randomly oriented PM vectors and do not 
have systemically younger ages. Since \citet{IRW07} 
could not find an age sequence of YSOs from NGC 6611, 
they argued that there is no significant sign of 
feedback-driven star formation in the vicinity of 
the southern pillars. Our results support the 
argument of their study. 

Therefore, we suggest that the formation of the 
stars seen in optical and infrared passbands was 
not triggered by the feedback from the massive 
stars in NGC 6611. The stars in Region 1 are likely 
to have formed spontaneously. \citet{MWN06} used 
smoothed particle hydrodynamics modeling to infer 
the physical states of the southern pillars. They 
also suggested that the shock front has not passed 
through the pillars yet and that these pillars 
are in a transition stage just before star formation. 
The southern pillars would still be possible sites 
of feedback-driven star formation.

The stars in the Region 2 
are systematically moving away from NGC 6611 
(see Figures~\ref{fig14} and ~\ref{fig15}). 
The fact that the RVs of these stars 
are similar to that of the northeastern 
pillar likely indicates a physical connection 
between them. This small system has a systemic 
RV of about 25 km s$^{-1}$, which is larger than 
the systemic RV of NGC 6611 (22.4 km s$^{-1}$). 
If we adopt $E(B-V) = 0.78$, four out of the five 
stars with RV measurements are brighter than 
the isochrone corresponding to 1 Myr at given 
colors, indicating they are younger than 1 Myr. 
These findings support that the formation of 
the stars in the vicinity of the northeastern 
pillar was induced by feedback from the 
massive stars in NGC 6611.

\section{Summary}\label{sec5}
We performed a kinematic study of young stellar 
population in M16, part of the Serpens OB1 
association. The results from this study are 
summarized as below. 

A total of 345 stars were selected as the members 
of M16 by analyzing the Gaia data \citep{gdr3}, optical 
and near-infrared spectra, the mid-infrared photometric 
data from the GLIMPS survey \citep{CBM09}, the list 
of X-ray sources \citep{BGP13}, and the database 
of Morgan-Keenan classification \citep{Sk14}. The spatial 
distribution of the members reveals that this SFR contains 
an open cluster (NGC 6611) and a distributed stellar 
population spread over $\sim 30$ pc. We also confirmed 
the presence of molecular clouds extending from north 
to east using the $^{13}$CO $J = 1 - 0$ line. This fact 
indicates that the massive stars in NGC 6611 have already 
blown out the southern clouds.

We determined the distance to M16 from the inversion 
of the Gaia parallaxes. The reddening toward M16 was 
estimated by analyzing the published $UBV$ photometric 
data of early-type members \citep{HJ56,HM69,HJI61}. 
The reddening values range from 0.78 to 1.32, indicating 
that there is differential reddening across this SFR. 
We investigated the color excess ratios in the optical 
and near-infrared bands and confirmed the abnormal 
reddening law ($R_{V,cl} = 4.5 \pm 0.1$). The ages 
of members, estimated by comparing the CMD 
with the theoretical isochrones, range 
from 1 Myr to 4 Myr.

The kinematics of members were investigated using 
the PMs from Gaia DR3 \citep{gdr3} and RVs obtained 
from spectroscopic observations. The vectorial
angle distribution of PMs shows the expansion of 
NGC 6611. This pattern of expansion is more pronounced 
over $r_h$. In addition, the cluster members have 
a symmetric RV distribution on an axis oriented 
northwest-southeast, which indicates rotation of 
this cluster. 

The distributed population do not have global 
patterns like radial expansion or stream motions 
in their PMs. We investigated the relationship 
between the gas pillars and the neighboring stars. 
Some stars in the Region 1 have similar RVs 
to those of the southern pillars, however their 
PM vectors are randomly oriented. These stars 
show a large age spread. On the other hand, the 
stars in the Region 2 are systematically receding 
away from NGC 6611 in PMs and their RVs are 
similar to those of the northeastern pillar. 
In addition, they belong to the youngest members 
in M16.

Based on these results, we discussed the star 
formation process in M16 from a 
kinematic perspective. Star formation probably 
occurred spontaneously throughout M16. In particular, 
the region with a high star formation efficiency might 
have formed NGC 6611. Some of the stars spread out in 
this SFR may have originated from the expansion of 
NGC 6611, but the proper motion vectors of these stars 
with random orientations do not strongly support 
this idea. Gas pillars located at the ridge of the 
H {\scriptsize II} bubble are possible sites of 
star formation by feedback from massive 
stars. The most likely site is the northeastern 
pillars in the Region 2; however, their contribution 
to the total stellar population is not very high.

We have performed a kinematic survey of the Galactic 
OB associations. A common finding of these systematic 
survey is that the formation process of stellar 
associations appears to be dominated by spontaneous 
star formation on timescales of millions of years. 
However, other models that explain the formation 
of stellar associations through cluster expansion 
and stellar feedback should not be ruled out. The 
final release of the Gaia data and forthcoming 
extensive spectroscopic surveys will lead to more 
definite conclusions about star formation process 
at spatial scales of tens of parsecs.

\section{acknowledgments}
The authors would like to thank the anonymous reviewers for 
their constructive comments. The authors would also like to 
express gratitude to Dr. Hyunwoo Kang for providing guidance 
on TRAO observations, and to Dr. Nelson Caldwell and the 
mountain staff for their assistance with Hectochelle observations. 
Observations reported here were conducted at the MMT Observatory, 
a joint facility of the University of Arizona and the Smithsonian Institution. This 
paper has made use of data obtained under the K-GMT Science 
Program (PIDs: MMT-2020A-001 and MMT-2021A-001) partly supported 
by the Korea Astronomy and Space Science Institute (KASI) 
grant funded by the Korean government (MSIT;No. 2023-1-860-02, 
International Optical Observatory Project) and from the European 
Space Agency (ESA) mission Gaia (https://www.cosmos.esa.int/gaia), 
processed by the Gaia Data Processing and Analysis Consortium 
(DPAC, https://www.cosmos.esa.int/web/gaia/dpac/consortium). Funding 
for the DPAC has been provided by national institutions, in particular
the institutions participating in the Gaia Multilateral Agreement.
The Digitized Sky Surveys were produced at the
Space Telescope Science Institute under U.S. Government
grant NAG W-2166. The images of these surveys are based on
photographic data obtained using the Oschin Schmidt Telescope 
on Palomar Mountain and the UK Schmidt Telescope.
The plates were processed into the present compressed digital
form with the permission of these institutions. 
This research has made use of the SIMBAD database, operated 
at CDS, Strasbourg, France and the data obtained with the Immersion Grating 
Infrared Spectrometer (IGRINS) that was developed 
under a collaboration between the University 
of Texas at Austin and the Korea Astronomy and 
Space Science Institute (KASI) with the financial 
support of the Mt. Cuba Astronomical Foundation, 
of the US National Science Foundation under 
grants AST-1229522 and AST-1702267, of the McDonald 
Observatory of the University of Texas at Austin, 
of the Korean GMT Project of KASI, and Gemini Observatory. 
The results of this work is based in part on observations made with the 
NASA/ESA/CSA JWST. The data were obtained from the Mikulski Archive 
for Space Telescopes at the Space Telescope Science Institute, 
which is operated by the Association of Universities for Research in 
Astronomy, Inc., under NASA contract NAS 5-03127 for JWST. 
These observations are associated with the program 2739.
The authors acknowledge PI Klaus Pontoppidan for developing 
their observing program with a zero-exclusive-access period. 
Based on observations made with the Harlan J. Smith Telescope, 
owned by the University of Texas at Austin at the McDonald Observatory, Texas.
Based on observations obtained at the international Gemini Observatory, a program of NSF NOIRLab, which is managed by the Association of Universities for Research in Astronomy (AURA) under a cooperative agreement with the U.S. National Science Foundation on behalf of the Gemini Observatory partnership: the U.S. National Science Foundation (United States), National Research Council (Canada), Agencia Nacional de Investigaci\'{o}n y Desarrollo (Chile), Ministerio de Ciencia, Tecnolog\'{i}a e Innovaci\'{o}n (Argentina), Minist\'{e}rio da Ci\^{e}ncia, Tecnologia, Inova\c{c}\~{o}es e Comunica\c{c}\~{o}es (Brazil), and Korea Astronomy and Space Science Institute (Republic of Korea). This work was supported by the National Research Foundation 
of Korea (NRF) grant funded by the Korean government (MSIT; 
grant Nos. RS-2022-NR072247 and 2022R1C1C2004102) and the 
research grant of Kongju National University in 2025. This 
research was partly supported by the Korea Astronomy and Space 
Science Institute under the R\&D program(Project No. 2025-1-860-02, 
Project No. 2025-1-868-02) supervised by the Korea AeroSpace Administration.
BL is grateful for Ms. Seulgi Kim’s assistance in data reduction 
and Prof. Jeong-Eun Lee's comments on observing proposal.

%

\vspace{5mm}
\facilities{MMT:6.5m, Gemini:South, TRAO:14m, Smith, JWST}


\software{{\tt xcsao} \citep{KM98}, {\tt NumPy} \citep{HMvdW20}, {\tt Scipy} \citep{VGO20}}




\bibliography{aas}{}
\bibliographystyle{aasjournal}



\end{document}